\documentclass[sigconf]{acmart}

\renewcommand\footnotetextcopyrightpermission[1]{} 
\usepackage{multirow}
\usepackage{graphicx}  
\usepackage{stfloats} 
\usepackage{enumitem} 
\usepackage{makecell}
\usepackage{xcolor}

\usepackage{longtable}
\usepackage{booktabs}
\usepackage{amsmath}
\usepackage{array}
\newcolumntype{P}[1]{>{\raggedright\arraybackslash}p{#1}}

\newcommand\cyan[1]{\textcolor{cyan}{#1}}

\setcopyright{none}

\begin{document}
\title{On the Factors in Quantum Software Quality}

\author{Jianjun Zhao}

\affiliation{%
  \institution{Kyushu University, Japan}
}
\email{zhao@ait.kyushu-u.ac.jp}

\renewcommand{\shortauthors}{J. Zhao}

\begin{abstract}
As quantum computing evolves, the demand for high-quality quantum software is increasing. Classical software quality models, such as McCall's model, provide a useful foundation, but they do not explicitly capture distinctive properties of quantum software, including probabilistic measurement outcomes, backend-dependent behavior, and hybrid quantum-classical workflows. This paper proposes a base quality model for quantum software using McCall's factor-criterion-metric structure as the organizing backbone. It broadens the set of quality factors by drawing on representative classical software quality models and standards and introduces quantum-specific extensions only where needed. We define the resulting factor set and quality criteria, provide a many-to-many mapping of core factor-to-criterion relationships, discuss representative candidate metrics together with the execution, reference, and interpretation conditions needed to use them, and illustrate how the resulting model can be applied across the quantum software lifecycle. The model is intended as a common starting point for structuring, assessing, and communicating quality concerns in quantum software while supporting context-specific adaptation. Further expert feedback, case studies, and empirical studies can be used to evaluate and refine the model and its application across different quantum software contexts.
\end{abstract}



\maketitle

\section{Introduction}
\label{sec:introduction}

Quantum software quality is becoming a central concern as quantum hardware, software stacks, and applications evolve~\cite{murillo2025roadmap}. Compared with classical software, quantum software has distinctive properties, constraints, and execution dependencies that complicate quality assessment and improvement. Measurement outcomes are generally probabilistic; behavior is sensitive to noise and device conditions; execution depends on specific backends and their calibration; and many applications are implemented as hybrid quantum-classical workflows. In addition, quantum software must account for quantum-specific physical properties and constraints, including the entanglement properties required for the intended computation and the no-cloning principle, which have no direct counterparts in classical systems. These properties, constraints, and dependencies make it difficult to assess and communicate software quality in quantum computing using classical assumptions alone.

In classical software engineering, software quality has long been supported by well-established factor models and standards. Representative models and standards provide established terminology and structures for relating high-level quality goals to quality criteria and measurable evidence. Early examples include the model proposed by McCall et al.~\cite{mccall1977factors}, commonly referred to as McCall's model, and Boehm's model~\cite{boehm1978characteristics}; later standards include ISO/IEC 9126-1:2001~\cite{iso9126-2001} and ISO/IEC 25010~\cite{iso25010-2011,iso25010-2023}. However, these classical models were not developed to explicitly address quantum-specific concerns such as probabilistic measurement outcomes, backend-dependent behavior, and hybrid quantum-classical workflows. As a result, to our knowledge, quantum software engineering still lacks a widely used base quality model that can serve as a common starting point for structuring, assessing, and communicating quality concerns in quantum software.

This paper proposes a base quality model for quantum software. We adopt McCall's factor-criterion-metric structure~\cite{mccall1977factors} as the organizing backbone, draw on representative later models and standards to broaden the classical factor set beyond McCall's original factors, and introduce quantum-specific extensions only when important concerns cannot be adequately represented through the retained classical concepts. We first preserve and compare the original factor terminology of the classical models and standards, and then use these sources to construct a consolidated factor set with definitions adapted to quantum software. We also make the construction logic explicit by stating the design principles and procedure used to build the model. The resulting model provides a set of quality criteria, a many-to-many mapping of core factor-to-criterion relationships, and representative candidate metrics for evaluating selected criteria under stated execution, reference, and interpretation conditions. It is designed to support context-specific tailoring rather than to prescribe a single fixed quality
profile.

The proposed base quality model can be tailored to specific application domains, platforms, organizational contexts, risk profiles, and evidence needs. The present criterion definitions, representative candidate metrics, and lifecycle illustrations are grounded primarily in current quantum software practice, especially backend-dependent, hybrid quantum-classical, and NISQ-era settings. They emphasize the settings in which assessment and measurement can currently be discussed most concretely. Practitioners can select and prioritize relevant factors and criteria and use metrics appropriate to their execution and tool settings to evaluate the selected criteria. The usefulness, completeness, and mappings of the model can be further evaluated and refined through expert feedback, case studies, and empirical studies.

This paper makes the following contributions:

\begin{itemize}
[leftmargin=15pt,itemsep=4pt,topsep=5pt,parsep=0pt]

\item We construct a base quality factor set for quantum software by consolidating the terminology of representative classical quality models and standards, retaining McCall's organizing structure, and introducing four quantum-specific factors where important quantum-specific concerns cannot be adequately represented by the retained classical factors. We also make explicit the design principles and construction procedure underlying this factor set (Section~\ref{sec:quality-factors}; Tables~\ref{table:classical-factor-models} and~\ref{table:qs-factors}).

\item We define a set of quality criteria for the proposed factors and provide a many-to-many mapping of core factor-to-criterion relationships, thereby making explicit how assessable criteria provide evidence for judging high-level quality objectives (Section~\ref{sec:criteria}; Tables~\ref{table:criteria-definitions} and~\ref{table:factor-criterion-mapping}).

\item We organize candidate metrics into broad metric types, provide representative candidate metrics for the proposed criteria, and explain the execution, reference, and interpretation conditions needed to use them (Section~\ref{sec:metrics}; Table~\ref{table:criteria-candidate-metrics}).

\item We describe how the base model can be tailored to a stated application context and illustrate how selected factors, criteria, and metrics can guide quality evidence collection and the reassessment of quality claims across the quantum software lifecycle (Sections~\ref{sec:quality-factors} and~\ref{sec:lifecycle}).

\end{itemize}

The rest of this paper is organized as follows. Section~\ref{sec:background} provides the minimal background and summarizes representative classical quality models. Section~\ref{sec:quality-factors} presents the proposed quality factors, the principles and procedure used to construct the base model, and the approach to context-specific tailoring. Section~\ref{sec:criteria} discusses the quality criteria, and Section~\ref{sec:metrics} develops the metric layer and discusses metric interpretation. Section~\ref{sec:lifecycle} describes how the proposed model can be used across the quantum software lifecycle. Section~\ref{sec:discussion-challenge} discusses open challenges. Section~\ref{sec:related-work} reviews related work, and Section~\ref{sec:conclusion} concludes the paper and outlines future directions.

\section{Background}
\label{sec:background}

This section provides the background required to support the proposed quality model. It briefly introduces the basic concepts of quantum computing directly relevant to software quality, then reviews the development of classical software quality models that form the foundation of our approach.

\subsection{Basic Concepts of Quantum Computing}
\label{subsec:qc-basics}

Quantum software differs from classical software because its behavior is shaped by algorithms and implementations, the physical principles of quantum mechanics, and the constraints of current hardware platforms. This subsection introduces a compact set of quantum-computing concepts that are needed to understand the quality factors, criteria, and later metric use discussed in this paper. The goal is not to provide a full introduction to quantum computing, but to explain the concepts that most directly affect software quality.
We use the following terms throughout the paper:

\begin{itemize}
[leftmargin=15pt,itemsep=4pt,topsep=5pt,parsep=0pt]

  \item \textit{Qubit and quantum state.} A quantum program manipulates quantum states stored in quantum bits (\textit{qubits}). Unlike a classical bit, a qubit is described by a state that cannot, in general, be directly inspected during computation without disturbing it. As a result, the internal state of a quantum program is usually inferred indirectly from measurement outcomes rather than observed directly.
  
  \item \textit{Superposition.} A qubit can exist in a superposition of basis states. This means that before measurement, a quantum state may encode multiple possible outcomes in a single mathematical state description. Because measurement resolves this state probabilistically, the same program input may lead to different observed outcomes across repeated runs.
  
  \item \textit{Entanglement.} Qubits can be correlated in ways that cannot be decomposed into independent local states. Many quantum algorithms use entanglement as a computational resource. In practice, the entanglement properties required for the intended computation may be degraded by noise, hardware imperfections, and implementation or compilation choices, which makes their preservation a relevant quality concern.
  
  \item \textit{Quantum gates and circuits.} Quantum computations are typically expressed as circuits composed of quantum gates. Ideally, gates implement \textit{unitary} transformations on the quantum state until measurement occurs. Circuit depth, gate count, and especially the number of two-qubit gates are commonly used as cost proxies because they affect both execution overhead and exposure to noise.
  
  \item \textit{Measurement and shots.} Measurement produces a classical outcome according to a probability distribution and generally changes the post-measurement state. When properties are estimated from measurement statistics, a circuit is typically executed repeatedly (\textit{shots}), and quantities such as outcome frequencies or expectation values are estimated from the resulting samples. These estimates are subject to finite-sampling uncertainty.
  
  \item \textit{Noise and decoherence.} Real quantum devices are affected by gate errors, readout errors, crosstalk, and decoherence. These effects introduce additional errors and variability in observed results beyond the intrinsic probabilistic nature of quantum measurement and can make results sensitive to device conditions. When an ideal target state, operation, or output distribution is available, appropriate fidelity, distance, or error measures can be used to quantify deviation from that target.
  
  \item \textit{Backend, calibration, and connectivity.} A \textit{backend} is a concrete execution target, such as a physical quantum device or a simulator. A physical backend has device-specific constraints, including limited qubit connectivity, native gate sets, and time-varying calibration parameters. Even when the logical circuit is unchanged, results may differ across backends or across different calibration periods on the same backend.
  
  \item \textit{Compilation/transpilation and mapping.} Before execution, a circuit is compiled or transpiled so that it satisfies backend constraints, for example by rewriting gates into the native gate set and mapping logical qubits and interactions onto physical qubits and device connectivity. Toolchains often apply multiple compilation passes over intermediate representations (IRs). This process can change depth, gate counts, and error sensitivity. In practice, compilation may also depend on user-specified settings and heuristic choices, such as optimization levels, pass choices, and transpilation seeds, which can introduce additional variability if not controlled.
  
  \item \textit{Ancilla, reversibility, and measurement placement.} Quantum algorithms and toolchains may introduce auxiliary qubits (\textit{ancilla}) and rely on reversible computation regions. Measurement placement is also important because measurement both produces classical information and changes the quantum state. Compilation and optimization should preserve declared assumptions about ancilla usage, reversibility regions, and where measurements occur.
  
  \item \textit{No-cloning principle.} An arbitrary unknown quantum state cannot be perfectly copied. This no-cloning principle constrains program abstractions, data flow, and interface design, and affects how quantum data can be passed, reused, and reasoned about.
  
 \item \textit{Error mitigation vs.\ error correction.} Error mitigation aims to reduce the effect of noise by post-processing or circuit-level techniques without requiring full fault tolerance. Error correction, in contrast, aims at long-term fault-tolerant computation through structured encoding and additional physical resources. Many current quantum software systems are evaluated under mitigation-oriented rather than fully fault-tolerant assumptions.
  
  \item \textit{NISQ regime.} Most currently available platforms operate in the noisy intermediate-scale quantum (NISQ) regime~\cite{preskill2018quantum}, where limited scale, noise, and hardware constraints strongly influence program behavior and the practical meaning of software quality claims.
  
  \item \textit{Parameterized circuits and VQAs.} Many applications use parameterized quantum circuits whose gate parameters are updated by a classical optimizer. A \textit{variational quantum algorithm (VQA)} iterates between quantum execution and classical optimization, and an \textit{ansatz} denotes the chosen parameterized circuit structure. In such settings, software quality depends not only on a fixed circuit, but also on the surrounding optimization loop and parameter-management process.
  
  \item \textit{Hybrid quantum-classical workflows and orchestration.} Quantum programs are often embedded in hybrid workflows, where classical software orchestrates circuit execution, parameter updates, result aggregation, and decision making. Quality depends on both the quantum component and the correctness and robustness of the surrounding classical orchestration, including measurement-conditioned control decisions.
\end{itemize}

Measurement plays a central role in quantum software quality. Because observing a quantum state yields a classical outcome drawn from a probability distribution, correctness and reliability are often assessed statistically rather than deterministically. This probabilistic nature complicates testing, debugging, and validation, and directly affects later discussions in this paper on correctness, reliability, measurement-related evidence, and reproducibility.

Quantum software is also strongly shaped by backend dependence, toolchain effects, and hybrid quantum-classical orchestration. Under NISQ assumptions~\cite{preskill2018quantum}, the same logical program may behave differently across backends, across time, or under different compilation settings, while many practical systems are embedded in workflows in which classical software controls execution, processes measurement results, and coordinates repeated runs or parameter updates. Backend and toolchain dependence, execution variability, and hybrid orchestration are directly related to later quality concerns such as portability, quantum hardware adaptability, reproducibility, orchestration correctness, and physical validity.

\subsection{Overview of Classical Software Quality Models}
\label{subsec:classical-quality-models}

Classical software quality models have long provided structured representations of software quality for describing, evaluating, and improving software systems~\cite{mccall1977factors,boehm1978characteristics}. They distinguish high-level quality concerns from the more specific assessable properties and measures used to evaluate those properties and provide evidence for judging the higher-level concerns, although individual models differ in their terminology, organization, and scope.
Table~\ref{table:classical-factor-models} summarizes representative classical models and standards and preserves the factor terminology used in their original sources. Items placed in the same row indicate related or corresponding concepts, but do not imply that their definitions are identical. The table is used to identify recurring quality concerns and terminological relationships across the classical literature, rather than to provide a complete historical survey.

\begin{table*}[th]
\centering
\caption{Comparison of Quality Factors in Classical Models and Standards}
\label{table:classical-factor-models}
\renewcommand\arraystretch{1.48}
\scriptsize
\begin{tabular}{|p{1.19cm}|p{1.4cm}|p{1.24cm}|p{1.24cm}|p{1.24cm}|p{1.14cm}|p{1.24cm}|p{1.24cm}|p{1.24cm}|p{1.24cm}|p{1.25cm}|}
\hline 
\multicolumn{1}{|p{1.2cm}|}{\bf McCall {\it et al.}~\cite{mccall1977factors}} &
\multicolumn{1}{p{1.25cm}|}{\bf Boehm {\it et al.}~\cite{boehm1978characteristics}} &
\multicolumn{1}{p{1.24cm}|}{\bf Murine and Carpenter~\cite{murine1983applying}} &
\multicolumn{1}{p{1.24cm}|}{\bf Bowen {\it et al.}~\cite{bowen1985specification}} &
\multicolumn{1}{p{1.24cm}|}{\bf Evans and Marciniak~\cite{evans1987software}} &
\multicolumn{1}{p{1.14cm}|}{\bf FURPS~\cite{grady1987software}} &
\multicolumn{1}{p{1.24cm}|}{\bf Deutsch and Willis~\cite{deutsch1988software}} &
\multicolumn{1}{p{1.24cm}|}{\bf Dromey~\cite{dromey1996cornering,dromey1995model}} &
\multicolumn{1}{p{1.24cm}|}{\bf ISO/IEC 9126-1: 2001~\cite{iso9126-2001}} &
\multicolumn{1}{p{1.24cm}|}{\bf ISO/IEC 25010: 2011~\cite{iso25010-2011}} &
\multicolumn{1}{p{1.24cm}|}{\bf ISO/IEC 25010: 2023~\cite{iso25010-2023}}\\
\hline

Correctness & & Correctness & Correctness & Correctness & & Correctness & & & & \\
\hline
& & & & & Functionality & & Functionality & Functionality & & \\
\hline
& & & & & & & & & Functional Suitability & Functional Suitability \\
\hline
Reliability & Reliability & Reliability & Reliability & Reliability & Reliability & Reliability & Reliability & Reliability & Reliability & Reliability \\
\hline
Efficiency & Efficiency & Efficiency & Efficiency & Efficiency & & Efficiency & Efficiency & Efficiency & & \\
\hline
& & & & & Performance & & & & & \\
\hline
& & & & & & & & & Performance Efficiency & Performance Efficiency \\
\hline
& Human Engineering & & & & & & & & & \\  
\hline
Usability & & Usability & Usability & Usability & Usability & Usability & Usability & Usability & Usability & \\  
\hline
& & & & & & & & & & Interaction Capability \\  
\hline
Integrity & & Integrity & Integrity & Integrity & & Integrity & & & & \\
\hline
& & & & & & & & & Security & Security \\
\hline
Maintainability & & Maintainability & Maintainability & Maintainability & & Maintainability & Maintainability & Maintainability & Maintainability & Maintainability \\
\hline
& Understandability & & & & & & & & & \\
\hline
& & & & & Supportability & & & & & \\
\hline
Flexibility & & Flexibility & Flexibility & Flexibility & & Flexibility & & & & Flexibility \\
\hline
& Modifiability & & & & & & & & & \\
\hline
Testability & Testability & Testability & & Testability & & & & & & \\
\hline
& & & Verifiability & & & Verifiability & & & & \\
\hline
Portability & Portability & Portability & Portability & Portability & & Portability & Portability & Portability & Portability & \\
\hline
Reusability & & Reusability & Reusability & Reusability & & Reusability & Reusability & & & \\
\hline
Interoperability & & Interoperability & Interoperability & Interoperability & & Interoperability & & & & \\
\hline
& & & & & & & & & Compatibility & Compatibility \\
\hline
& & & Survivability & & & Survivability & & & & \\
\hline
& & & Expandability & & & Expandability & & & & \\
\hline
& & & & Documentation & & & & & & \\
\hline
& & & & & & Safety & & & & Safety\\
\hline
& & Intraoperability & & & & & & & & \\
\hline
& & & & & & Manageability & & & &\\
\hline
& & & & & & & Process-mature & & &\\
\hline
\end{tabular}

\vspace{1mm}
\begin{minipage}{0.99\linewidth}
\raggedright
\scriptsize
\textit{Note.} Items placed in the same row represent related or corresponding concepts across models; they do not necessarily have identical definitions.
\end{minipage}

\end{table*}

McCall's model is particularly relevant to this paper because it organizes software quality through a factor-criterion-metric structure. This structure distinguishes high-level quality objectives from assessable properties and concrete measures. Later models and standards broaden the range of quality concerns represented in classical quality models by introducing or emphasizing additional concerns such as compatibility, security, supportability, interaction capability, and safety. The specific principles and procedures used to select and consolidate these concerns into the proposed factor set, and to extend that factor set for the quantum context, are presented in Section~\ref{sec:quality-factors}.

These classical models provide an important foundation, but they were not developed to explicitly represent several quantum-specific concerns that become central to software quality assessment in the quantum context.

\subsection{Limitations of Classical Models in the Quantum Context}
\label{subsec:limitations}

Classical software quality models have been extensively studied and applied in classical software engineering, but they do not fully address quantum-specific quality concerns arising from quantum computation and its execution context. The reason is not that classical quality concerns become irrelevant in the quantum setting. On the contrary, many classical concerns, such as correctness, reliability, maintainability, portability, and usability, remain important. The problem is that the assumptions under which these concerns are typically interpreted in classical software do not fully match the behavior and execution context of quantum software.

\begin{itemize}
[leftmargin=15pt,itemsep=4pt,topsep=5pt,parsep=0pt]

\item \emph{Classical models do not explicitly capture several quantum-specific concerns that directly affect software behavior and quality assessment.} Behavioral assessment of quantum software often relies on probabilistic measurement outcomes rather than direct observation of its full internal quantum state, so assessments of \emph{correctness} and \emph{reliability} often depend on statistical evidence under explicitly stated execution conditions. In addition, quantum behavior may depend strongly on the entanglement properties required for the intended computation, measurement placement, backend noise, calibration drift, and compilation or transpilation choices. Classical models generally do not make such concerns explicit, nor do they distinguish clearly between properties that can be judged directly from software artifacts and those that require assessment under stated execution settings.

\item \textit{Classical quality models can account for platform and environment dependence, but they do not make the rapidly changing backend, calibration, and toolchain context of quantum software a central part of quality assessment.} The same logical program may show different behavior across devices, simulators, calibration periods, transpilation settings, or dependency versions. As a result, quality concerns such as \emph{portability} and \emph{reproducibility} become more context-sensitive, and \emph{quantum hardware adaptability} requires explicit treatment. These issues are not absent from classical software engineering, but they become much more central in quantum software.

\item Quantum software is often developed and executed as part of hybrid quantum-classical workflows, where quality depends on the quantum components and their interaction with classical control logic, data exchange, synchronization, result interpretation, and execution orchestration. Classical quality models include general concerns such as interoperability and integration, but they do not explicitly represent the quantum-classical execution boundary or the orchestration- and synchronization-specific evidence needed to assess it.

\item Some important concerns in quantum software are tied to physical validity rather than only to conventional software behavior. For example, software abstractions and transformations may need to preserve intended unitary and measurement semantics, remain consistent with physical constraints such as the no-cloning principle, and maintain stated physical or semantic assumptions throughout compilation and execution. A software artifact or transformation may pass ordinary toolchain checks while no longer satisfying the conditions under which its behavior is intended to be interpreted. Classical quality models do not directly provide concepts for making such concerns explicit.
\end{itemize}

These limitations motivate the proposed base quality model for quantum software. Our goal is not to discard classical quality models, but to retain their useful structures and quality concepts where they remain applicable, while introducing quantum-specific factors and criteria when those classical concepts are insufficient.

\section{Quality Factors for Quantum Software}
\label{sec:quality-factors}

This section defines quality factors for quantum software. The proposed model is a base model that can be tailored to different application domains, platforms, and organizational contexts. The model is rooted in McCall's factor model~\cite{mccall1977factors}, broadened by later classical models and standards, and further extended with a small set of quantum-specific factors for which classical concepts are no longer sufficient.

\subsection{Terminology and the Factor-Criterion-Metric View}
\label{subsec:terminology}

This subsection defines the terminology used throughout the paper. Quantum software refers to quantum programs (e.g., circuits), the classical control software that orchestrates quantum execution, and the associated artifacts used in development and operation, such as configurations, scripts, and documentation. Quality denotes the degree to which quantum software fulfills stated and implied requirements in a specified development and operational context. Because observed behavior in quantum software may vary with the execution setting, quality judgments in this context are inherently context-dependent.

Within this paper, a factor is a high-level quality objective or stakeholder-oriented quality concern. A factor expresses what kind of quality is sought, such as correctness, reliability, portability, or physical validity. A factor is not itself a single measurable quantity or a specific evaluation procedure. Instead, factors are assessed through criteria that identify more concrete properties to be examined when judging them.

A criterion is an assessable property used to assess one or more factors. It is more concrete than a factor, but it is not itself a specific measure or measurement procedure. Rather, a criterion identifies what property of the software, its artifacts, or its execution should be examined when assessing a factor. The same criterion may support more than one factor, depending on the quality concern being assessed.

A metric is a concrete measure used to evaluate a criterion under explicitly stated assumptions, evidence sources, execution settings, and measurement procedures. Candidate metrics may be expressed as counts, ratios, rates, distances, latencies, pass rates, coverage values, checklist-based scores, binary indicators, or other explicitly defined measures.

In this sense, factors represent high-level quality concerns, criteria are assessable properties through which those concerns are assessed, and metrics specify concrete measures for evaluating the criteria. In the quantum context, many metrics cannot be interpreted independently of the execution and measurement conditions under which they are obtained, including the backend, transpilation options, shot count, random seeds, and mitigation choices, or, where applicable, the reference or acceptance condition against which they are evaluated.

\begin{table*}[ht]
\centering
\caption{Definitions of Quantum Software Quality Factors}
\label{table:qs-factors}
\renewcommand\arraystretch{1.18}
\footnotesize
\begin{tabular}{|p{1.3cm}|p{2.8cm}|p{9.4cm}|p{2.6cm}|}
\hline
\textbf{Type} & \textbf{Factor} & \textbf{Definition} & \textbf{Related Terms in Classical Models and Standards} \\
\hline

\multirow{14}{*}{\centering\makecell[c]{\textbf{Classical}\\\textbf{(McCall)}}}
& \multicolumn{3}{l|}{\textbf{Product operation}} \\
\cline{2-4}
& Correctness & Extent to which quantum software satisfies its specified requirements under stated execution assumptions. &  \\
\cline{2-4}
& Reliability &
Extent to which quantum software performs its specified functions within stated acceptance conditions under specified execution conditions over a stated period or set of executions, without unacceptable interruptions or failures. &
\\
\cline{2-4}
& Efficiency & Extent to which quantum software achieves the required performance relative to the quantum and classical computational, storage, and orchestration resources used under stated execution conditions and device constraints. &
\makecell[l]{Performance (FURPS);\\Performance Efficiency\\(ISO/IEC 25010:2011, 2023)} \\
\cline{2-4}
& Integrity & Extent to which quantum software and related artifacts are protected against unauthorized access or modification. &  \\
\cline{2-4}
& Usability & Effort required to learn to use and operate quantum software, prepare its inputs, and interpret its outputs. &
\makecell[l]{Human Engineering\\(Boehm {\it et al.})} \\
\cline{2-4}

& \multicolumn{3}{l|}{\textbf{Product revision}} \\
\cline{2-4}
& Maintainability & Effort required to locate faults and to repair or modify quantum software. &  \\
\cline{2-4}
& Flexibility & Effort required to modify quantum software for use in other applications or environments. &  \\
\cline{2-4}
& Testability & Effort required to test quantum software and establish confidence in its behavior under controlled execution conditions. &  \\
\cline{2-4}

& \multicolumn{3}{l|}{\textbf{Product transition}} \\
\cline{2-4}
& Portability &
Effort required to transfer quantum software from one software, hardware, or execution environment to another, including changes in quantum backend or toolchain. &
\\
\cline{2-4}
& Reusability & Extent to which parts of quantum software can be reused in other applications with limited modification. &  \\
\cline{2-4}
& Interoperability &
Extent to which quantum software can exchange information with other systems or components and use the information that is exchanged in its intended environment. &
\\
\hline

\multirow{13}{*}{\centering\makecell[c]{\textbf{Classical}\\\textbf{(Ext.)}}}
& Functionality & Extent to which quantum software provides the functions required for its intended use. &
\makecell[l]{Functional suitability\\(ISO/IEC 25010:2011, 2023)} \\
\cline{2-4}
& Understandability & Extent to which the structure, purpose, and assumptions of quantum software can be readily understood. &  \\
\cline{2-4}
& Supportability &
Extent to which quantum software can be effectively operated, diagnosed, maintained, and evolved with the available support resources. & \\
\cline{2-4}
& Modifiability &
Extent to which quantum software can be effectively and efficiently modified without introducing defects or degrading its existing quality. &
\\
\cline{2-4}
& Compatibility &
Extent to which quantum software can, as required by its intended context, exchange information with other products or components and/or perform its required functions while sharing a common execution environment and resources. &
\\
\cline{2-4}
& Expandability &
Extent to which the functional scope, supported problem sizes, or data-handling capacity of quantum software can be extended without extensive restructuring. & \\
\cline{2-4}
& Survivability &
Extent to which quantum software can continue to provide its essential functions, or restore them within an acceptable time, in the presence of failures, attacks, or other disruptions. & \\
\cline{2-4}
& Safety & Extent to which quantum software avoids unacceptable harm or unsafe outcomes in its intended context of use. &  \\
\cline{2-4}
& Security & Extent to which quantum software, related assets, and execution workflows are protected against unauthorized access, misuse, and attack. &  \\
\cline{2-4}
& Verifiability & Extent to which properties of quantum software can be justified under explicit assumptions through formal, semi-formal, or empirical evidence. &  \\
\cline{2-4}
& Interaction Capability &
Extent to which specified users can interact with quantum software through its user interfaces to exchange information with the system and complete intended tasks. &
\\
\cline{2-4}
& Manageability &
Extent to which quantum software executions and experiments can be effectively configured, monitored, and controlled. & \\
\cline{2-4}
& Documentation &
Extent to which the documentation and related information artifacts for quantum software are complete, accurate, consistent, and usable for its correct use, assessment, maintenance, and evolution. &  \\
\hline

\multirow{4}{*}{\centering\textbf{Quantum}}
& Entanglement Robustness &
Extent to which the entanglement properties required for the intended computation are preserved under relevant physical and execution variations while satisfying stated acceptance conditions. &
\\
\cline{2-4}
& Quantum Hardware Adaptability &
Extent to which quantum software continues to satisfy stated acceptance conditions across quantum devices and under changes in their capabilities, constraints, and operating conditions. &
\\
\cline{2-4}
& Hybrid Quantum-Classical Interoperability &
Extent to which quantum and classical components can exchange and use information and jointly perform their required roles within hybrid workflows. &
\\
\cline{2-4}
& Physical Validity & Extent to which quantum software respects the physical constraints and execution semantics required for valid interpretation of its behavior. &  \\
\hline

\end{tabular}
\end{table*}

\subsection{Design Principles for Constructing the Base Model}
\label{subsec:design-principles}

The proposed base model is not constructed by simply collecting quality terms from earlier literature. Instead, it is guided by four principles intended to keep the model coherent and to provide a clear basis for later tailoring and refinement.

\begin{enumerate}
[leftmargin=15pt,itemsep=3pt,topsep=5pt,parsep=0pt]

\item \textit{The model maintains continuity with classical software quality research.} McCall's factor-criterion-metric structure is adopted as the organizing backbone, while later classical models and standards are used to identify recurring quality concerns that are incorporated into the model as additional classical factors when they are not explicitly named or sufficiently differentiated in McCall's original factor set.

\item \textit{The model is extended conservatively.} A quantum-specific extension is introduced only when an important concern in quantum software cannot be adequately expressed through the retained classical factors and criteria. The purpose is to preserve the classical foundation wherever possible rather than to replace it with an entirely separate set of quantum-specific factors and criteria.

\item \textit{The model maintains a clear distinction among factors, criteria, and metrics.} Factors represent high-level quality objectives, criteria represent assessable properties through which those objectives are assessed, and metrics are concrete measures used to evaluate criteria under stated assumptions. A quantum-specific concern is therefore introduced as a factor only when it represents a sufficiently stable and distinct high-level quality objective; otherwise, it is treated as a criterion supporting one or more factors.

\item \textit{The proposed model is intended to serve as a base model for context-specific adaptation.} Different application domains, platforms, and organizational contexts may require different subsets of factors, criteria, and metrics. The model provides a common structure that can be tailored and further refined through practical and empirical use.
\end{enumerate}

\subsection{Construction Procedure for the Base Model}
\label{subsec:construction-procedure}

The principles above are applied through the following construction procedure. Making this procedure explicit clarifies how the factor set, quality criteria, mappings, and representative candidate metrics are constructed and provides a common logic for later extension, comparison, tailoring, and empirical refinement. The resulting model remains open to later extension and refinement.

\begin{itemize}
[leftmargin=15pt,itemsep=4pt,topsep=5pt,parsep=0pt]

\item The first step is to identify and document the classical source terminology. Representative classical models and standards are examined, and their original factor terminology is preserved in Table~\ref{table:classical-factor-models}. McCall's model provides the primary organizing structure, while later models and standards identify additional recurring quality concerns.

\item The second step is to consolidate the relevant classical quality concepts into a coherent factor set. Closely related concepts are consolidated under canonical factor names in Table~\ref{table:qs-factors}. Related source terms are recorded only as cross-references and do not imply that their original definitions are strictly equivalent.

\item The third step is to identify concerns that are not adequately represented by the classical backbone. Such concerns are added as quantum-specific extensions only when their meaning cannot be expressed clearly and consistently through an appropriate quantum interpretation of the retained classical factors.

\item The fourth step is to assign each quantum-specific concern to the appropriate layer. Concerns representing high-level quality objectives are placed at the factor layer, whereas more specific assessable properties are placed at the criterion layer. The inherited layer assignments of the classical backbone are retained rather than
reclassified.

\item The fifth step is to define the criteria used to assess the resulting factors, establish the core factor-to-criterion mappings, and identify representative candidate metrics. Because many measurements of quantum software depend on the evaluation context in which they are obtained, the model does not impose a single universal metric set.
\end{itemize}

The resulting base model can then be adapted to a particular application domain, platform, or organizational context, as discussed in Section~\ref{subsec:tailorable-base-model}. This procedure makes the construction decisions explicit while allowing later additions and empirical refinements to be explained using the same general logic.

\subsection{Classical Backbone of the Proposed Model}
\label{subsec:classical-base}

The classical backbone of the proposed model is constructed from the representative factor models and standards summarized in Table~\ref{table:classical-factor-models}. McCall's factor model provides the primary organizing structure, while later models and standards inform the extension of this structure by identifying recurring quality concerns that are incorporated as additional classical factors when they are not explicitly named or sufficiently differentiated in McCall's original factor set.

We retain McCall's three groups because they provide a useful distinction among qualities related to product operation, product revision, and product transition. Product operation includes \emph{correctness}, \emph{reliability}, \emph{efficiency}, \emph{integrity}, and \emph{usability}. Product revision includes \emph{maintainability}, \emph{flexibility}, and \emph{testability}, while product transition includes \emph{portability}, \emph{reusability}, and \emph{interoperability}. The original grouping is retained for organizational continuity, although the interpretation of these factors is adapted to the quantum software context.

In particular, \emph{correctness} and \emph{reliability} often require statistical evidence under explicitly stated execution conditions; \emph{efficiency} includes both quantum and classical resources; \emph{portability} must account for backend and toolchain differences; and \emph{interoperability} must include the interfaces between quantum execution and classical control. These adaptations preserve the general meaning of the classical factors, while making their application conditions explicit for quantum software. The same conservative adaptation applies to the other retained classical factors: their general meanings are preserved, while their assessment scope is extended where necessary to include quantum programs, hybrid control software, execution configurations, toolchain conditions, and associated evidence artifacts.

Later models and standards contribute additional factors that broaden the classical backbone, including concerns related to \emph{functionality}, \emph{understandability}, \emph{supportability}, \emph{modifiability}, \emph{compatibility}, \emph{expandability}, \emph{survivability}, \emph{safety}, \emph{security}, \emph{verifiability}, \emph{interaction capability}, \emph{manageability}, and \emph{documentation}. These factors are included when they represent recurring high-level quality concerns that are useful for assessing quantum software but are not explicitly named or sufficiently differentiated in McCall's original factor set.

Table~\ref{table:qs-factors} presents the complete factor set, including the classical backbone and the quantum-specific extensions, together with definitions adapted to quantum software. For traceability, Table~\ref{table:classical-factor-models} preserves the terminology used in the original sources, while related source terms in Table~\ref{table:qs-factors} are provided only as a cross-reference and do not imply strict equivalence among their original definitions.

Consolidating factors from different quality models does not require merging every pair of related factors. We retain factors separately when they represent distinct high-level assessment targets, even if some of their supporting criteria or assessment evidence overlap. For example, \emph{functionality} concerns whether the required capabilities are provided, whereas \emph{correctness} concerns whether the implemented behavior conforms to its stated requirements or intended results. The factor \emph{integrity} retains the narrower concern of protecting software and data against unauthorized access or modification, whereas \emph{security} covers a broader set of concerns, including confidentiality, authenticity, and vulnerability resistance. Similarly, \emph{maintainability} concerns the overall effort required to locate faults, repair the software, and perform changes needed for its continued upkeep; \emph{modifiability} focuses more specifically on the ease of implementing changes in response to revised requirements, algorithms, hardware constraints, or operating conditions; and \emph{flexibility} concerns adapting the software for different applications, uses, or environments.

Other related factors are separated for the same reason. The factor \emph{usability} addresses the overall ability of users to learn, operate, and use the software effectively, whereas \emph{interaction capability} focuses more specifically on the quality of human-system interaction. The factor \emph{compatibility}, adopted from ISO/IEC 25010~\cite{iso25010-2011,iso25010-2023}, covers both co-existence in a shared environment and the ability to exchange information with other products or components, whereas the retained \emph{interoperability} factor focuses specifically on the exchange and use of information across systems; \emph{hybrid quantum-classical interoperability} further specializes that concern to the quantum-classical execution boundary. Finally, \emph{portability} concerns transferring software across environments or platforms, whereas \emph{quantum hardware adaptability} concerns maintaining acceptable behavior under backend-specific constraints and hardware variation. These distinctions preserve the meaning of the source models while making the assessment targets of the consolidated factor set explicit.

\subsection{Quantum-Specific Factors}
\label{subsec:quantum-extension}

The conservative-extension principle raises an important question: which quantum-specific concerns should be represented as factors rather than as criteria? We retain four quantum-specific factors because each represents a high-level assessment target that is not adequately captured by the retained classical factors alone and is supported by criteria that make its assessment scope explicit. Although these factors are related to classical quality concerns, they address quantum-specific quality objectives that require explicit representation in the factor set.

By contrast, measurement-related concerns such as measurement bias, statistical repeatability, and mitigation sensitivity are retained at the criterion layer. Evaluation of these criteria can provide evidence for several factors, including \emph{correctness}, \emph{reliability}, and \emph{verifiability}, rather than defining separate high-level quality factors.

\begin{itemize}
[leftmargin=15pt,itemsep=4pt,topsep=5pt,parsep=0pt]

\item \emph{Entanglement Robustness.}
For quantum software whose intended computation depends on entanglement, assessing overall reliability alone does not show whether the required entanglement properties are preserved. This factor isolates the preservation of those properties under relevant physical and execution variations. It applies when entanglement is an essential resource for the intended computation and is not a universal requirement for all quantum software.

\item \emph{Quantum Hardware Adaptability.}
The retained portability factor concerns the effort required to transfer software between environments. Quantum hardware adaptability concerns whether quantum software continues to satisfy stated acceptance conditions across quantum devices and as device capabilities, constraints, and operating conditions change. It is therefore treated as a separate assessment target from transfer effort itself.

\item \emph{Hybrid Quantum-Classical Interoperability.}
General interoperability concerns the exchange and use of information among software systems and components. Hybrid quantum-classical interoperability specializes this high-level concern to whether quantum and classical components can exchange and use information and jointly perform their required roles within hybrid workflows. Its supporting criteria capture the specific orchestration, synchronization, and robustness properties that provide evidence for that objective.

\item \emph{Physical Validity.}
Physical validity concerns whether quantum software respects the physical constraints and execution semantics required for its behavior to be interpreted as the intended quantum computation. A software artifact or transformation may pass conventional toolchain checks while relying on invalid quantum-data semantics, such as requiring perfect duplication of an arbitrary unknown quantum state, or while failing to preserve intended unitary or measurement semantics or other stated physical and semantic assumptions. Physical validity is therefore assessed as a distinct quality objective supported by evidence of unitarity preservation, measurement semantics soundness, no-cloning compliance, and assumption preservation.
\end{itemize}

\subsection{Tailoring the Base Model}
\label{subsec:tailorable-base-model}

The proposed base model is intended to support context-specific adaptation rather than to serve as a fixed quality profile. Tailoring begins with an explicit context profile describing the conditions under which quality is to be assessed. Depending on the application, this profile may specify the target domain and backends, execution mode, hardware and toolchain assumptions, risk profile, required quantum-classical workflow, stakeholder and organizational goals, and the expected level of reproducibility or formal assurance.

Based on this context profile, users select and prioritize the factors and criteria relevant to the intended quality claims. The core factor-to-criterion mappings in Table~\ref{table:factor-criterion-mapping} provide the default starting point. When a criterion has an additional relationship with a factor only under a particular algorithm, requirement, application domain, or operational assumption, that relationship may be introduced as a context-dependent mapping in the tailored quality profile. Candidate metrics are then selected for the retained criteria, together with the evidence requirements, reference conditions, uncertainty treatment, and acceptance thresholds needed to interpret the measurements. The resulting selection of factors, criteria, mappings, metrics, and associated assessment conditions constitutes a tailored quality profile, hereafter referred to simply as a quality profile.

A tailored quality profile should document the retained factors and criteria, the core mappings used and any context-dependent mappings added, the selected metrics, the assumptions and evidence context, and the acceptance conditions and reassessment triggers. The canonical names and definitions of the base model should be preserved where possible, while any addition, omission, or reinterpretation of a factor, criterion, or mapping should be explicitly recorded and justified. This documentation supports traceability of quality claims, comparison across tailored quality profiles, and continued connection to the common structure of the base model.

\section{Quality Criteria for Quantum Software}
\label{sec:criteria}

Building on the factor set defined in Section~\ref{sec:quality-factors}, this section summarizes quality criteria for quantum software. We focus on criteria as assessable properties used to assess factors. In quantum software, many classical criteria remain relevant, but their interpretation must reflect probabilistic measurement outcomes, backend-dependent behavior, and hybrid quantum-classical workflows.

\subsection{Overview of Quality Criteria}

Quality criteria are assessable properties used to assess software quality factors. Table~\ref{table:criteria-definitions} defines the set of quality criteria used in the proposed model. It contains criteria rooted in McCall's model, additional classical criteria needed to support the extended classical factor set, and criteria introduced to address quantum-specific concerns.

The additional classical criteria are included only when two conditions hold: (1) an extended classical factor in Table~\ref{table:qs-factors} cannot be adequately assessed using McCall's original set of quality criteria alone, and (2) the missing concern remains a general software quality concern rather than one arising specifically from quantum mechanics. Some of these criteria are aligned with notions made explicit in later standards, some adapt related second-level notions from later models, and others are synthesized in this paper to make recurring quality concerns explicit. These construction paths are not intended as mutually exclusive provenance categories, because later models and standards often use different names and levels of granularity for related concerns. The definitions in Table~\ref{table:criteria-definitions} are therefore the definitions adopted in the proposed model rather than verbatim reproductions of definitions from individual sources. Factor-level conceptual lineage is recorded in Tables~\ref{table:classical-factor-models} and~\ref{table:qs-factors}, while each additional criterion is retained only when it is needed to support an interpretable assessment of one or more factors in the proposed mapping. The additional criteria are intended as representative extensions rather than an exhaustive list.

When these criteria are applied to quantum software, they should be evaluated under explicitly stated development and operational conditions. Relevant conditions may include the backend, transpilation options, shot count, random seeds, mitigation settings, toolchain versions, and other assumptions that affect the collected evidence.

\begin{table*}[th]
\centering
\caption{Criteria Definitions Used in the Proposed Quality Model}
\label{table:criteria-definitions}
\renewcommand\arraystretch{1.23}
\scriptsize
\begin{tabular}{|p{1.1cm}|p{2.3cm}|p{13.1cm}|}
\hline
\multicolumn{1}{|l|}{\bf Type} &
\multicolumn{1}{l|}{\bf Criterion} &
\multicolumn{1}{l|}{\bf Definition} \\
\hline

\multirow{23}{*}{\centering\bf \makecell[c]{Classical\\ (McCall)}}
& Traceability & Those attributes of the quantum software that enable requirements or expected properties to be linked to relevant implementation artifacts, evaluation evidence, and execution conditions. \\
\cline{2-3}
& Completeness & Those attributes of the quantum software that determine whether all required functions, behaviors, and workflow elements are implemented for the stated operating context. \\
\cline{2-3}
& Consistency & Those attributes of the quantum software that determine whether design, representations, configurations, and implementation conventions are used uniformly and without contradiction. \\
\cline{2-3}
& Accuracy & Those attributes of the quantum software that determine whether calculations and outputs agree with stated reference values, distributions, or acceptance bounds under specified conditions. \\
\cline{2-3}
& Error tolerance & Those attributes of the quantum software that determine whether acceptable behavior can be maintained, possibly through controlled degradation, under stated non-nominal conditions. \\
\cline{2-3}
& Simplicity & Those attributes of the quantum software that avoid unnecessary structural, orchestration, and configuration complexity while preserving required functionality. \\
\cline{2-3}
& Modularity & Those attributes of the quantum software that organize quantum and classical components into cohesive units with limited dependencies and well-defined interfaces. \\
\cline{2-3}
& Generality & Those attributes of the quantum software that determine the range of input instances, parameter settings, and deployment contexts supported without rewriting core logic. \\
\cline{2-3}
& Expandability & Those attributes of the quantum software that enable supported functions, problem sizes, or data-handling capacity to be extended without extensive restructuring. \\
\cline{2-3}
& Instrumentation & Those attributes of the quantum software that enable relevant information about executions, configurations, intermediate artifacts, and outcomes to be observed and recorded for monitoring, diagnosis, and assessment. \\
\cline{2-3}
& Self-descriptiveness & Those attributes of the quantum software that make the purpose, assumptions, and interpretation of its artifacts explicit within or directly associated with those artifacts. \\
\cline{2-3}
& Execution efficiency & Those attributes of the quantum software that determine whether processing time meets stated requirements under specified execution conditions. \\
\cline{2-3}
& Storage efficiency & Those attributes of the quantum software that determine whether storage use meets stated requirements without compromising required auditability or reproducibility. \\
\cline{2-3}
& Access control & Those attributes of the quantum software that enforce specified authorization policies for access to software artifacts, data, credentials, execution services, and result repositories. \\
\cline{2-3}
& Access audit & Those attributes of the quantum software that enable access to and modification of software, configurations, data, and result artifacts to be recorded and traced. \\
\cline{2-3}
& Operability & Those attributes of the quantum software that enable intended users to initiate, monitor, control, and complete its intended workflows under stated operating conditions. \\
\cline{2-3}
& Training & Those attributes of the quantum software that support intended users in learning to configure, operate, and interpret the software correctly. \\
\cline{2-3}
& Communicativeness & Those attributes of the quantum software that enable intended users to understand required inputs, execution status, outputs, and the basic meaning of probabilistic results. \\
\cline{2-3}
& \makecell[l]{Software system\\independence} &
Those attributes of the quantum software that minimize unnecessary dependence on particular operating systems, runtime services, libraries, and toolchains. \\
\cline{2-3}
& Machine independence & Those attributes of the quantum software that determine the extent to which it avoids unnecessary dependence on particular classical host hardware systems and general hardware interfaces. \\
\cline{2-3}
& \makecell[l]{Communications\\commonality} &
Those attributes of the quantum software that enable communication through standard or agreed protocols and interface mechanisms. \\
\cline{2-3}
& Data commonality & Those attributes of the quantum software that use standard or agreed representations for input data, circuit artifacts, configurations, metadata, and results. \\
\cline{2-3}
& Conciseness & Those attributes of the quantum software that support implementation of its intended functions without unnecessary code or redundancy, while preserving clarity and explicit assumptions. \\
\hline

\multirow{17}{*}{\centering\bf \makecell[c]{Classical\\(Ext.)}} & Resource utilization & Those attributes of the quantum software that determine whether the amounts and types of quantum and classical computational resources used while performing its functions meet stated requirements. \\ \cline{2-3} 
& Readability & Those attributes of the quantum software that make source code, circuit representations, configuration artifacts, and documentation easy for intended readers to understand under stated quantum assumptions. \\
\cline{2-3}
& Change locality & Those attributes of the quantum software that allow changes to be made locally with limited ripple effects across circuits, settings, and orchestration logic. \\
\cline{2-3}
& Documentation adequacy & Those attributes of the quantum software that provide sufficient documentation for correct use, reproduction, and interpretation, including settings and evidence rules. \\
\cline{2-3}
& Tool support & Those attributes of the quantum software that provide the interfaces, artifacts, metadata, and automation hooks needed to support effective testing, debugging, analysis, maintenance, and operation. \\ \cline{2-3}
& Configurability & Those attributes of the quantum software that support controlled and auditable configuration of parameters and execution settings without ad hoc code changes. \\
\cline{2-3}
& Reproducibility &
Those attributes of the quantum software that enable its workflow to be re-executed from recorded artifacts, dependencies, configurations, and procedures, yielding results or conclusions consistent with the original within stated tolerances. \\
\cline{2-3}
& Recoverability & Those attributes of the quantum software that enable restoration or resumption of acceptable operation after a failure or interruption under stated recovery policies. \\
\cline{2-3}
& Confidentiality
& Those attributes of the quantum software that protect sensitive assets (e.g., credentials, circuits, configurations, data, and results) from unauthorized disclosure. \\
\cline{2-3}
& Authenticity & Those attributes of the quantum software that enable the identities and origins of relevant entities and artifacts to be verified. \\
\cline{2-3}
& Vulnerability resistance & Those attributes of the quantum software that reduce exploitable weaknesses and resist misuse and attacks across the hybrid workflow boundary. \\
\cline{2-3}
& Hazard control & Those attributes of the quantum software that identify and control hazards and unacceptable risks in safety- or security-critical contexts. \\
\cline{2-3}
& Fail-safe behavior & Those attributes of the quantum software that ensure that invalid conditions or failures lead to controlled termination, safe fallback, or another specified safe state rather than misleading or uncontrolled behavior. \\
\cline{2-3}
& Interface conformance & Those attributes of the quantum software that determine whether interfaces conform to specified contracts for APIs, data formats, and semantics in hybrid workflows. \\
\cline{2-3}
& Version compatibility &
Those attributes of the quantum software that enable its required functions and interfaces to remain usable across specified versions of relevant toolchains, libraries, platforms, and data formats without unintended behavioral changes or extensive adaptation.  \\
\cline{2-3}
& Result transparency &
Those attributes of the quantum software that make reported results interpretable by clearly reporting the execution conditions, uncertainty, aggregation, mitigation, and post-processing steps used to produce them.  \\
\cline{2-3}
& Co-existence &
Those attributes of the quantum software that enable it to perform its required functions while sharing a common execution environment and resources with other products or components without adverse interference. \\
\hline

\end{tabular}
\end{table*}

\begin{table*}[th]\ContinuedFloat
\centering
\caption{Criteria Definitions Used in the Proposed Quality Model (continued)}
\renewcommand\arraystretch{1.4}
\scriptsize
\begin{tabular}{|p{1.1cm}|p{2.3cm}|p{13.1cm}|}
\hline
\multicolumn{1}{|l|}{\bf Type} &
\multicolumn{1}{l|}{\bf Criterion} &
\multicolumn{1}{l|}{\bf Definition} \\
\hline

\multirow{16}{*}{\centering\bf Quantum}
& Entanglement stability & Those attributes of the quantum software that determine whether the entanglement properties required for the intended computation remain within stated acceptance bounds in the presence of noise and decoherence. \\
\cline{2-3}
& Entanglement backend robustness &
Those attributes of the quantum software that determine whether the entanglement properties required for the intended computation remain within stated acceptance bounds across specified quantum backends under comparable execution conditions. \\
\cline{2-3}
& Entanglement compilation robustness &
Those attributes of the quantum software that determine whether the entanglement properties required for the intended computation remain within stated acceptance bounds when compilation and transpilation choices are varied under controlled conditions. \\
\cline{2-3}
& Backend portability &
Those attributes of the quantum software that enable it to be transferred to and executed on specified quantum backends with limited adaptation while continuing to satisfy stated behavioral requirements. \\
\cline{2-3}
& Calibration robustness &
Those attributes of the quantum software that determine whether it continues to satisfy stated acceptance conditions as calibration and other time-varying device conditions change.  \\
\cline{2-3}
& Constraint tolerance &
Those attributes of the quantum software that determine whether it can satisfy stated acceptance conditions when executed under specified quantum-device constraints, such as connectivity, native gate sets, and available-qubit limits.  \\
\cline{2-3}
& Orchestration correctness &
Those attributes of the quantum software that determine whether the sequencing and coordination of quantum and classical components conform to the specified control-flow, data-flow, component-invocation, and result-handling requirements of the hybrid workflow. \\ 
\cline{2-3} 
& Synchronization correctness &
Those attributes of the quantum software that determine whether specified ordering, timing, and dependency constraints among quantum execution events, the availability of measurement results, and classical actions are satisfied during hybrid-workflow execution. \\
\cline{2-3}
& Orchestration robustness &
Those attributes of the quantum software that enable a hybrid workflow to continue satisfying stated acceptance conditions, recover within specified bounds, or enter a defined safe or controlled-degradation state when orchestration-related disruptions occur. \\
\cline{2-3}
& Measurement bias & 
Those attributes of the quantum software that determine whether estimates derived from measurement outcomes exhibit systematic measurement-related deviations from stated reference values or distributions under nominally fixed conditions, beyond expected sampling variation. \\ \cline{2-3}
& Statistical repeatability & 
Those attributes of the quantum software that determine whether empirical outcome distributions, estimated quantities, or derived decisions remain statistically consistent across repeated executions under nominally identical conditions.\\
\cline{2-3}
& Mitigation sensitivity &
Those attributes of the quantum software that determine the extent to which results or conclusions vary when specified mitigation configurations and associated post-processing choices are changed while other relevant conditions are held fixed. \\
\cline{2-3}
& Unitarity preservation &
Those attributes of the quantum software that determine whether regions intended to denote unitary transformations continue to denote unitary transformations after representation, optimization, compilation, or other program transformations. \\
\cline{2-3}
& \makecell[l]{Measurement\\semantics soundness} & 
Those attributes of the quantum software that determine whether the measurement basis and placement, outcome interpretation, post-measurement state evolution, and measurement-conditioned classical control conform to the specified semantics. \\
\cline{2-3}
& No-cloning compliance &
Those attributes of the quantum software that determine whether its abstractions, interfaces, and transformations avoid operations or semantics that require arbitrary unknown quantum states to be perfectly duplicated. \\
\cline{2-3}
& Assumption preservation &
Those attributes of the quantum software that determine whether the declared physical and semantic assumptions required for the intended computation remain valid after toolchain transformations and during execution under stated conditions. \\
\hline
\end{tabular}
\end{table*}

\subsection{Factor-to-Criterion Mapping}
\label{subsec:factor-criterion-mapping}

Table~\ref{table:factor-criterion-mapping} presents the core mappings between the factors in Table~\ref{table:qs-factors} and the criteria used to assess them. A mapping is included when the criterion has a stable conceptual relationship with the factor and, whenever that criterion is applicable to the software being assessed, its evaluation can provide direct and interpretable evidence for the factor without requiring an additional algorithm-, domain-, or project-specific argument to establish the relationship. The mapping is many-to-many: a factor is normally assessed through several criteria, and the same criterion may support more than one factor. The mapping should therefore not be interpreted as a strict decomposition in which each criterion belongs exclusively to one factor, nor does a favorable evaluation of one mapped criterion alone establish the corresponding factor.

Some criteria may also have secondary or context-dependent relationships with other factors. Such relationships are not included in the core mapping when they depend on a particular algorithm, requirement, application domain, or operational assumption. They may instead be added and justified in a tailored quality profile.

The core mappings shown in the table are intended as a practical base rather than a fixed or exhaustive assignment. A tailored quality profile may select only the criteria relevant to its prioritized factors and may introduce additional criteria or mappings when required by a particular domain, organizational context, or operational setting. However, any addition, omission, or reinterpretation should be documented so that the relationship between the quality claim and its supporting evidence remains explicit.

\begin{table*}[t]
\centering
\caption{Core Factor-to-Criterion Mappings
in the Quantum Context}

\label{table:factor-criterion-mapping}
\renewcommand\arraystretch{1.4}
\footnotesize
\begin{tabular}{|p{1.6cm}|p{3.0cm}|p{11.7cm}|}
\hline
\textbf{Type} & \textbf{Factor} & \textbf{Criteria} \\
\hline

\multirow{14}{*}{\centering\makecell[c]{\textbf{Classical}\\\textbf{(McCall)}}}
& \textbf{Product operation} &  \\
\cline{2-3}
& Correctness & Traceability, Completeness, Consistency, Accuracy, Measurement bias, Orchestration correctness \\
\cline{2-3}
& Reliability & Accuracy, Error tolerance, Consistency, Simplicity, Measurement bias, Statistical repeatability, Mitigation sensitivity, Calibration robustness, Orchestration robustness \\
\cline{2-3}
& Efficiency & Execution efficiency, Storage efficiency, Resource utilization \\
\cline{2-3}
& Integrity & Access control, Access audit \\
\cline{2-3}
& Usability & Operability, Training, Communicativeness, Documentation adequacy, Result transparency \\
\cline{2-3}

& \textbf{Product revision} &  \\
\cline{2-3}
& Maintainability & Consistency, Simplicity, Modularity, Conciseness, Self-descriptiveness, Readability, Change locality, Documentation adequacy, Tool support \\
\cline{2-3}
& Flexibility & Generality, Expandability, Modularity, Self-descriptiveness \\
\cline{2-3}
& Testability & Instrumentation, Tool support, Modularity, Simplicity, Self-descriptiveness \\
\cline{2-3}

& \textbf{Product transition} &  \\
\cline{2-3}
& Portability & Software system independence, Machine independence, Backend portability, Version compatibility, Modularity, Self-descriptiveness \\
\cline{2-3}
& Reusability & Generality, Modularity, Self-descriptiveness, Software system independence, Machine independence \\
\cline{2-3}
& Interoperability &
Communications commonality, Data commonality, Interface conformance \\
\hline

\multirow{13}{*}{\centering\makecell[c]{\textbf{Classical}\\\textbf{(Ext.)}}}
& Functionality & Traceability, Completeness, Consistency \\
\cline{2-3}
& Understandability & Simplicity, Self-descriptiveness, Consistency, Readability \\
\cline{2-3}
& Supportability & Training, Operability, Communicativeness, Self-descriptiveness, Documentation adequacy, Tool support \\
\cline{2-3}
& Modifiability & Modularity, Consistency, Simplicity, Conciseness, Self-descriptiveness, Change locality \\
\cline{2-3}
& Compatibility &
Co-existence, Communications commonality, Data commonality, Interface conformance, Version compatibility \\
\cline{2-3}
& Expandability & Expandability, Modularity, Generality, Change locality, Configurability \\
\cline{2-3}
& Survivability & Error tolerance, Instrumentation, Operability, Recoverability, Orchestration robustness \\
\cline{2-3}
& Safety & Error tolerance, Traceability, Consistency, Operability, Hazard control, Fail-safe behavior \\
\cline{2-3}
& Security & Access control, Access audit, Confidentiality, Authenticity, Vulnerability resistance \\
\cline{2-3}
& Verifiability & Traceability, Consistency, Self-descriptiveness, Instrumentation, Reproducibility, Result transparency, Mitigation sensitivity, Assumption preservation \\
\cline{2-3}
& Interaction Capability & Communicativeness, Operability, Training, Result transparency \\
\cline{2-3}
& Manageability & Instrumentation, Operability, Self-descriptiveness, Consistency, Configurability, Traceability \\
\cline{2-3}
& Documentation & Self-descriptiveness, Traceability, Training, Documentation adequacy, Result transparency \\
\hline

\multirow{4}{*}{\centering\makecell[c]{\textbf{Quantum}}}
& Entanglement Robustness & Entanglement stability, Entanglement backend robustness, Entanglement compilation robustness \\
\cline{2-3}
& Quantum Hardware Adaptability & Backend portability, Calibration robustness, Constraint tolerance \\
\cline{2-3}
& Hybrid Quantum-Classical Interoperability & Orchestration correctness, Synchronization correctness, Orchestration robustness \\
\cline{2-3}
& Physical Validity & Unitarity preservation, Measurement semantics soundness, No-cloning compliance, Assumption preservation \\
\hline

\end{tabular}
\end{table*}

\subsection{Classical Criteria in the Quantum Context}

Classical factor-based models, such as McCall's model~\cite{mccall1977factors}, provide an established factor-criterion structure and a set of classical quality criteria. Using the criterion definitions in Table~\ref{table:criteria-definitions} as reference, we provide brief usage notes for classical criteria when assessing quantum software. We first cover criteria rooted in McCall's model, and then present the additional classical criteria introduced in this paper to support the extended classical factor set in Table~\ref{table:qs-factors}. In the quantum context, criteria should be evaluated under explicitly stated execution settings, as outcomes and failures can be probabilistic and backend-dependent. For criteria rooted in McCall's model, we retain McCall's original criterion terminology to preserve consistency with the source model.

\subsubsection{Classical Criteria Rooted in McCall's Model}

The criteria in this subsection retain the terminology of McCall's original model, while their interpretation is adapted to the characteristics and execution conditions of quantum software. The following discussion highlights the aspects that require particular attention in the quantum context rather than repeating the definitions given in Table~\ref{table:criteria-definitions}.

\begin{itemize}
[leftmargin=15pt,itemsep=4pt,topsep=5pt,parsep=0pt]

\item \textit{Traceability.}
For quantum software, traceability should connect requirements or expected properties to both the implementation and the concrete execution context used for evaluation. This context should include relevant settings such as backend, transpilation options, shot count, random seeds, and mitigation choices, because these can materially affect outcomes.

\item \textit{Completeness.}
In quantum software, completeness should be judged with respect to the intended operating context, including required hybrid workflows and supported execution modes such as simulation and real-device runs.

\item \textit{Consistency.}
In quantum workflows, consistency also includes uniform conventions for representing and interpreting probabilistic evidence, together with consistent configuration and implementation conventions such as qubit ordering, parameter naming, seed recording, and output formats. It concerns uniformity of design, representation, and implementation rather than statistical agreement across repeated executions, which is addressed by statistical repeatability.

\item \textit{Accuracy.}
For quantum software, accuracy is typically evaluated statistically under fixed settings. It should therefore be stated with respect to an explicit reference value or distribution, an acceptance condition, or an error bound, together with the measurement or estimation procedure used. It supports correctness by indicating whether outputs agree with stated reference values, distributions, or acceptance bounds, and reliability when such accuracy is maintained under stated execution conditions.

\item \textit{Error tolerance.}
In quantum settings, non-nominal conditions include increased noise, transient backend instability, time-varying calibration, and failures in execution services. Error tolerance, therefore, concerns whether the software continues to provide acceptable behavior under such conditions, as defined by the stated operational assumptions. When continued operation is not possible, entry into a controlled safe state is addressed separately through \emph{fail-safe behavior}.

\item \textit{Simplicity.} 
For quantum software, simplicity concerns both circuit structure and the surrounding orchestration and configuration logic, because unnecessary complexity in either can hinder review, diagnosis, and controlled experimentation. Raw resource counts, such as circuit depth and two-qubit gate count, should be interpreted primarily as resource-utilization measures unless they are normalized or combined with structural complexity measures.

\item \textit{Modularity.} 
For quantum software, modularity concerns the extent to which quantum circuits, classical components, and hybrid workflow elements are organized into cohesive units with limited dependencies and well-defined interfaces. Module count alone does not indicate good modularity; evaluation should instead consider coupling, cohesion, interface complexity, and the extent to which changes can be confined to individual modules.

\item \textit{Generality.}
For quantum software, generality often means supporting a range of input instances, parameter regimes, and deployment contexts without rewriting core logic, while making explicit the assumptions under which such breadth holds.

\item \textit{Expandability.}
As a criterion, expandability concerns the internal software attributes that allow computational functions, data-handling capacity, or supported problem sizes to be extended without extensive restructuring. It supports the assessment of the broader expandability factor, together with modularity, generality, change locality, and configurability.

\item \textit{Instrumentation.}
For quantum software, instrumentation commonly includes logging run metadata and intermediate artifacts such as compiled circuits, transpilation reports, and configuration snapshots, because post hoc diagnosis is otherwise difficult under limited observability and probabilistic measurement outcomes.

\item \textit{Self-descriptiveness.}
In quantum software, self-descriptiveness should make explicit key assumptions and interpretation rules, such as what counts as success, how results are aggregated, and which settings are fixed or variable.

\item \textit{Execution efficiency.}
For quantum software, execution efficiency should be stated with respect to the measured components and experimental conditions, since total processing time may include compilation or transpilation, queue and execution latency, and classical orchestration overhead.

\item \textit{Storage efficiency.}
In quantum experimentation, storage may include run logs, raw measurement outcomes, provenance records, and intermediate artifacts. Storage efficiency, therefore, concerns keeping these artifacts manageable without undermining auditability and reproducibility.

\item \textit{Access control.}
For quantum software, access control includes execution credentials, configuration files, experiment scripts, and result artifacts, since unauthorized changes to any of these can invalidate conclusions.

\item \textit{Access audit.}
In quantum workflows, access audit is important for preserving traceability and integrity of empirical evidence, because changes to configurations, backends, or result artifacts can materially affect reported outcomes.

\item \textit{Operability.}
For quantum software, operability concerns whether intended workflows can be initiated, monitored, controlled, and completed by their intended users under stated operating conditions. The software should provide clear operational procedures and status information for quantum execution and the surrounding classical workflow. Repeatability, recovery from failures, and robustness under disruptions are addressed separately through their corresponding criteria.

\item \textit{Training.}
In quantum software, training depends strongly on documentation quality, runnable examples, and clear guidance on configuration and interpretation, because incorrect assumptions about settings or statistical evidence can easily lead to misuse.

\item \textit{Communicativeness.}
For quantum software, communicativeness concerns whether inputs, outputs, execution status, and the basic meaning of probabilistic results are presented in a form that intended users can understand and use. Detailed reporting of uncertainty, run settings, mitigation, aggregation, and post-processing is addressed separately by \emph{result transparency}.

\item \textit{Software system independence.}
For quantum software, software system independence concerns the extent to which workflows avoid unnecessary dependence on particular classical software environments, runtime services, libraries, or toolchains. Recording such dependencies supports reproducibility, but reproducibility itself is evaluated separately by whether the workflow can be re-executed from recorded artifacts, dependencies, configurations, and procedures and yields results or conclusions consistent with the original within stated tolerances.

\item \textit{Machine independence.}
In quantum software, machine independence concerns dependency on the underlying classical host systems and general hardware interfaces. Portability across quantum execution backends is addressed separately by backend portability.

\item \textit{Communications commonality.}
For hybrid quantum-classical workflows, communications commonality concerns whether components use common or standardized communication protocols and mechanisms. Conformance to a particular API, data format, or semantic contract is addressed separately through \emph{interface conformance}.

\item \textit{Data commonality.}
In quantum workflows, data commonality includes standardized representations for circuits, configurations, metadata, and results, enabling reuse, interoperability, and consistent post-processing.

\item \textit{Conciseness.}
For quantum software, conciseness should be balanced with explicit assumptions and interpretability of evidence, because overly compressed implementations may obscure critical experimental conditions or semantic constraints.
\end{itemize}

\subsubsection{Additional Classical Criteria Beyond McCall’s Model}

Table~\ref{table:criteria-definitions} defines the additional classical criteria introduced to support the extended classical factors. Rather than restating each definition, this subsection highlights the main distinctions that are important when these criteria are applied to quantum software.

\emph{Resource utilization} concerns the amounts and types of quantum and classical resources used, including qubits, circuit depth, critical gate counts, shot count, and classical processing and orchestration resources. It is distinct from \emph{execution efficiency}, which concerns processing time, and \emph{storage efficiency}, which concerns storage requirements. \emph{Readability} also requires quantum-specific assumptions, such as measurement placement, qubit ordering, and statistical interpretation rules, to be made explicit. \emph{Change locality} concerns whether modifications to circuits, transpilation settings, orchestration logic, or related dependencies remain localized rather than producing unnecessary non-local effects. \emph{Documentation adequacy} requires sufficient documentation of algorithm intent, execution settings, mitigation and post-processing steps, and the way evidence is aggregated and judged. It concerns external documentation, whereas \emph{self-descriptiveness} concerns explanatory information contained in or closely associated with the software artifacts themselves.

\emph{Tool support} concerns whether the software exposes the interfaces, artifacts, metadata, and automation hooks needed for effective testing, debugging, analysis, maintenance, and operation; it does not merely refer to whether external tools exist. \emph{Configurability} concerns controlled changes to backends, transpilation options, shot count, random seeds, and mitigation settings without ad hoc code modifications. \emph{Reproducibility} concerns whether a workflow can be re-executed from recorded artifacts, dependencies, configurations, and procedures, yielding results or conclusions consistent with the original within stated tolerances. Relevant sources of variability, such as the backend, execution period, transpilation settings and random seeds, shot count, and mitigation choices, should be recorded together with an indication of which variations are expected to affect the results. Reproducibility differs from \emph{statistical repeatability}, which concerns agreement across repeated executions under nominally identical conditions. \emph{Recoverability} concerns restoring or resuming acceptable operation after job failures, retries, partial results, or other disruptions under stated re-execution policies, while preserving the traceability of the resulting evidence. It differs from \emph{orchestration robustness}, which concerns the broader behavior of a hybrid workflow under disruptions, including recovery, fallback, and controlled degradation.

The security- and safety-related criteria extend across the complete hybrid workflow. \emph{Confidentiality} concerns the protection of credentials, circuits, configurations, datasets, and results; \emph{authenticity} concerns whether the identities and origins of compiled circuits, configuration files, result artifacts, and other relevant entities can be verified. Unauthorized modification is addressed separately through \emph{integrity}, while \emph{vulnerability resistance} concerns exploitable weaknesses across APIs, orchestration scripts, cloud services, and data pipelines. In safety- or security-critical workflows, \emph{hazard control} requires unacceptable risks to be identified and controlled through conservative assumptions, explicit validation conditions, and clearly stated operational boundaries. \emph{Fail-safe behavior} concerns preventing silent misuse of settings, reporting invalid configurations clearly, and entering a controlled state rather than producing misleading conclusions from insufficient or non-comparable evidence.

\emph{Interface conformance} concerns adherence to agreed APIs, data formats, and semantic contracts. In hybrid workflows, it should also prevent semantic mismatches among circuit execution, result parsing, and downstream classical logic, thereby supporting both interoperability and compatibility. \emph{Version compatibility} concerns whether required functions and interfaces remain usable across specified toolchain and platform versions without unintended behavioral changes or extensive adaptation. Because compiler and transpiler changes may materially affect generated circuits and outcomes, dependency pinning and compatibility information form part of the evidence context. \emph{Result transparency} requires the settings, uncertainty, mitigation, aggregation, and post-processing used to produce a result to be reported clearly. It is more specific than \emph{communicativeness}, which concerns whether inputs and outputs can be understood and used by their intended users.

\subsection{Quantum-Specific Criteria}

The quantum-specific criteria used in this model fall into two groups. The first group primarily supports the assessment of the four quantum-specific factors defined in Section~\ref{subsec:quantum-extension}. They concern the preservation of required entanglement properties, adaptation to quantum hardware, coordination between quantum and classical components in hybrid workflows, and compliance with the physical and semantic constraints of quantum computation. Some of them may also support related classical factors, as reflected in the core factor-to-criterion mappings in Table~\ref{table:factor-criterion-mapping}. The second group contains measurement- and mitigation-related criteria arising from quantum measurement, probabilistic measurement outcomes, and associated post-processing practices. These criteria remain at the criterion layer because they support the assessment of several classical factors, particularly \emph{correctness}, \emph{reliability}, and \emph{verifiability}, rather than defining separate high-level quality factors.
Table~\ref{table:criteria-definitions} provides the definitions of these criteria. The following discussion focuses on their practical interpretation and their role in the factor-criterion mapping.

\subsubsection{Criteria Supporting the Quantum-Specific Factors}

These criteria capture the main assessable properties used to assess the four quantum-specific factors introduced in Section~\ref{subsec:quantum-extension}.

\begin{itemize}
[leftmargin=15pt,itemsep=2pt,topsep=4pt,parsep=0pt]

\item \emph{Entanglement Stability.}
In practice, entanglement stability should be evaluated under clearly stated execution settings, because entanglement-related evidence can be sensitive to noise and decoherence. Evaluation may compare an entanglement measure or witness with a stated target or baseline, or quantify its variation across repeated executions under nominally identical conditions.

\item \emph{Entanglement Backend Robustness.}
This criterion is used to judge whether the entanglement properties required for the intended computation remain within stated acceptance bounds across specified quantum backends under comparable execution conditions, and whether conclusions that depend on those properties remain supported.

\item \emph{Entanglement Compilation Robustness.}
This criterion is used to judge whether the entanglement properties required for the intended computation remain within stated acceptance bounds when compilation or transpilation passes, parameters, and optimization settings are varied under controlled conditions, and whether conclusions that depend on those properties remain supported.

\item \emph{Backend Portability.}
Backend portability concerns whether quantum software can be transferred to and executed on specified quantum backends with limited adaptation while continuing to satisfy stated behavioral requirements. Evaluation should consider both the adaptation required and the behavior obtained after transfer. The criterion supports the assessment of the classical factor \emph{portability} as well as \emph{quantum hardware adaptability}; it does not imply independence from backend-specific gate sets, connectivity, compilation constraints, or execution conditions.

\item \emph{Calibration Robustness.}
Calibration robustness is judged by whether the software continues to satisfy stated acceptance conditions as calibration and other time-varying device conditions change under otherwise comparable execution conditions.

\item \emph{Constraint Tolerance.}
Constraint tolerance is judged by whether the software satisfies stated acceptance conditions when compilation and execution are subject to specified quantum-device constraints, such as connectivity, native gate sets, available-qubit limits, circuit-depth limits, and error budgets.

\item \emph{Orchestration Correctness.}
This criterion concerns whether the sequencing and coordination of quantum and classical components conform to the specified control-flow, data-flow, component-invocation, parameter-passing, and result-handling requirements of the hybrid workflow. It evaluates the workflow logic rather than the correctness of the quantum circuit in isolation.

\item \emph{Synchronization Correctness.}
This criterion concerns whether specified ordering, timing, and dependency constraints among quantum execution events, the availability of measurement results, and classical actions are satisfied during hybrid-workflow execution. For example, a measurement-conditioned classical action must not be performed before the required measurement result becomes available.

\item \emph{Orchestration Robustness.}
This criterion concerns whether a hybrid workflow continues to satisfy stated acceptance conditions, recovers within specified bounds, or enters a defined safe or controlled-degradation state when orchestration-related disruptions occur. It distinguishes successful recovery and controlled degradation from unhandled or uncontrolled failure.

\item \emph{Unitarity Preservation.}
Unitarity preservation concerns whether a region intended to denote a unitary transformation continues to denote a unitary transformation after representation and toolchain transformations. The criterion concerns the semantics of the region as a whole, rather than merely whether each syntactic operation is a unitary gate. It applies only to regions intended to be unitary; measurement, reset, and other explicitly non-unitary operations may legitimately appear elsewhere in the computation.

\item \emph{Measurement Semantics Soundness.}
This criterion concerns whether transformations preserve the intended measurement basis and placement, outcome interpretation, post-measurement state evolution, and measurement-conditioned classical control. It is particularly important when later operations or decisions depend on measurement results.

\item \emph{No-Cloning Compliance.}
This criterion concerns whether quantum-data abstractions, interfaces, and program transformations avoid operations or semantics that require arbitrary unknown quantum states to be perfectly duplicated. It is particularly relevant to high-level quantum data abstractions, compiler transformations, and distributed quantum workflows, where software-level representations or data-handling rules must remain consistent with the no-cloning constraint. Copying classical information obtained from measurement, or preparing multiple systems in the same known state, does not constitute a violation of this criterion.

\item \emph{Assumption Preservation.}
Assumption preservation is judged by whether the declared physical and semantic assumptions required for the intended computation remain valid after toolchain transformations and during execution under stated conditions.
\end{itemize}

\subsubsection{Measurement-Related Criteria Supporting Classical Factors}

These criteria address measurement- and mitigation-related properties of quantum execution that provide evidence for assessing classical quality factors, particularly correctness, reliability, and verifiability.

\begin{itemize}
[leftmargin=15pt,itemsep=4pt,topsep=5pt,parsep=0pt]

\item \emph{Measurement Bias.}
This criterion concerns persistent measurement-related deviation from a stated reference after accounting for expected finite-sampling variation. It should therefore be evaluated with respect to an explicit reference condition and a stated statistical estimation procedure. Quantum software may be statistically repeatable but still biased if it produces stable yet systematically shifted results; conversely, run-to-run variation alone does not establish measurement bias. It primarily supports \emph{correctness}, and secondarily \emph{reliability}, because quality claims are commonly inferred from measurement outcomes.

\item \emph{Statistical Repeatability.}
This criterion concerns whether empirical outcome distributions, estimated quantities, or derived decisions remain statistically consistent across repeated executions under nominally identical conditions, while allowing for expected sampling variation. It primarily supports \emph{reliability}. Unlike \emph{reproducibility}, which concerns whether a workflow can be re-executed from recorded artifacts and conditions, statistical repeatability concerns repeated execution under nominally identical conditions.

\item \emph{Mitigation Sensitivity.}
This criterion concerns the extent to which results or conclusions vary when specified mitigation configurations and associated post-processing choices are changed while other relevant conditions are held fixed. It supports \emph{reliability} and \emph{verifiability}, because excessive dependence on a particular mitigation configuration can weaken both the stability of conclusions and the ability to assess them independently.

\end{itemize}

\section{Candidate Metrics for Quantum Software Quality}
\label{sec:metrics}

This section develops the metric layer of the proposed quality model. It first explains the role and required elements of candidate metrics, then organizes relevant metric types and presents representative candidate metrics for the quality criteria defined in Section~\ref{sec:criteria}. The candidate metrics are intended to be instantiated within a tailored quality profile under explicitly stated execution, reference, and interpretation conditions.

\subsection{Role of Metrics in the Factor-Criterion-Metric Structure}
\label{subsec:metrics-role}

Within the factor-criterion-metric structure of the proposed model, factors represent high-level quality concerns, criteria are assessable properties through which those concerns are assessed, and metrics are concrete measures used to evaluate the criteria. Their role is to provide measurable evidence at the criterion layer rather than to serve as factor-level scores.

The metric layer follows the structural discipline of classical software quality models, in which criteria are supported by measures that can be collected, computed, checked, interpreted, and, where necessary, repeated under controlled conditions. Classical software quality metrics commonly include ratios, checklist scores, and structural counts over requirements, designs, source code, documentation, and test artifacts. Quantum software also involves these artifacts, but its assessment additionally depends on circuits, transpilation results, measurement outcomes, shot budgets, backend properties, calibration data, noise behavior, error mitigation procedures, and hybrid optimization traces. The metric layer therefore also includes numerical error metrics, distributional metrics, statistical uncertainty measures, resource metrics, robustness metrics, and cross-execution agreement measures.

In this paper, the metrics are presented as candidate metrics. This means that they are concrete enough to identify a measurement object and a measurement function or counting rule, but they are not proposed as standardized quality measures. A candidate metric becomes operational only when it is instantiated in a quality profile with the required evaluation context and, where applicable, reference condition, aggregation rule, and acceptance threshold. This position is important because many quantum software metrics are meaningful only relative to a specified backend, simulator, noise model, shot count, transpilation policy, optimizer setting, or calibration window. Therefore, a metric value should not be interpreted in isolation from the context in which it is obtained.

\subsection{Measurement Objects and Evidence Sources}
\label{subsec:measurement-objects}

A metric requires a measurement object. For classical software, common measurement objects include requirements specifications, design descriptions, source code, documentation, test plans, test results, problem reports, and operational records. These objects remain relevant for quantum software, especially for hybrid quantum-classical programs, quantum libraries, verification artifacts, and reproducibility packages. However, quantum software introduces additional measurement objects that are central to quality assessment.

The main measurement objects in the quantum software context include:

\begin{itemize}
[leftmargin=15pt,itemsep=4pt,topsep=5pt,parsep=0pt]

\item Problem specifications and domain assumptions, such as Hamiltonians, oracle definitions, input distributions, and physical or mathematical assumptions;
\item Quantum algorithms and high-level quantum program representations;
\item Source-level quantum-classical programs;
\item Logical circuits and intermediate circuit representations;
\item Transpiled circuits and mapping results;
\item Backend configurations, coupling maps, basis gates, calibration records, and noise models;
\item Measurement outcomes, output distributions, expectation values, confidence intervals, and repeated-execution records;
\item Error mitigation configurations and mitigation reports;
\item Optimizer traces and convergence records in hybrid algorithms;
\item Documentation, notebooks, scripts, random seeds, environment files, and other reproducibility artifacts.
\end{itemize}

These objects provide different forms of evidence. Some metrics can be collected statically from artifacts, such as circuit depth, two-qubit gate count, dependency on backend-specific gates, or documentation completeness. Some candidate metrics require controlled execution, such as success probability, value error, distributional distance, between-run variation, or mitigation-induced shift. Other candidate metrics require comparison across execution contexts, such as cross-backend agreement, transpilation-induced variation, variation across calibration windows, or shot-budget sensitivity. For this reason, candidate metrics should always be interpreted together with the evidence sources from which they are obtained.

\subsection{Required Elements of a Candidate Metric}
\label{subsec:metric-elements}

A candidate metric in this paper is not merely a label or a general measurement direction. It is a quantitative or checkable measure proposed for evaluating a quality criterion. For operational use within the factor-criterion-metric structure, an instantiated candidate metric should make explicit what is measured, how it is measured, under what conditions it is measured, and how the resulting value should be interpreted. 

In the quantum software context, this discipline is particularly important because a metric value may change with execution settings, backend properties, transpilation choices, noise behavior, shot budgets, calibration data, and mitigation procedures. For example, an observed success probability, an energy error, or a distributional distance is not meaningful unless the execution context and reference condition are stated. Similarly, a structural metric such as circuit depth or two-qubit gate count should be interpreted together with the circuit representation, optimization level, target backend, and gate set used to obtain it. Therefore, each candidate metric should be defined with enough information to support its later instantiation, repeated measurement, comparison, and interpretation.

An operational instantiation of a candidate metric should specify the following elements.

\begin{itemize}
[leftmargin=15pt,itemsep=3pt,topsep=4pt,parsep=0pt]

    \item \textit{Criterion.} The quality criterion that the metric is intended to evaluate. A metric may provide evidence for more than one criterion, but its interpretation should be stated separately for each criterion.

    \item \textit{Measurement object.} The artifact, behavior, or record being measured, such as a requirement specification, source program, quantum circuit, transpiled circuit, backend execution record, measurement distribution, optimizer trace, mitigation report, or reproducibility package.

    \item \textit{Required input data.} The data needed to compute or check the metric, such as the number of requirements, traced artifacts, circuit statistics, shot outcomes, reference values, backend configurations, calibration records, or repeated-run results.

    \item \textit{Measurement function or counting rule.} The formula, ratio, checklist rule, binary rule, statistical estimator, distance function, or counting procedure used to obtain the metric value. This element distinguishes a metric from a vague measurement idea.

    \item \textit{Evaluation context.} The artifact and execution context in which the metric is obtained, including the lifecycle phase, artifact version, simulator or backend, shot count, random-seed policy, transpilation policy, optimization level, noise model, mitigation setting, calibration window, and other relevant conditions.

    \item \textit{Interpretation rule.} The direction and meaning of the metric value, such as whether smaller or larger values indicate better quality, whether the value should be compared with a threshold, and whether it should be interpreted as an error, coverage, agreement, stability, cost, or risk indicator.

    \item \textit{Reference or acceptance condition.} Whether the metric requires a reference value, reference distribution, oracle, expected property, agreement condition, or acceptance bound. Some metrics are reference-based, such as value error or distributional distance, whereas others are reference-free, such as circuit depth, documentation coverage, or between-run variation.

    \item \textit{Aggregation rule.} The way metric values are combined when they are collected over multiple modules, circuits, test cases, backends, executions, or lifecycle artifacts. Aggregation may use a mean, maximum, minimum, weighted score, pass rate, checklist score, or profile-specific rule. Statistical uncertainty summaries, such as confidence intervals, should be reported separately when applicable.
\end{itemize}

These elements do not imply that every metric must have the same mathematical form. Some metrics are ratios, such as the number of traced quality claims divided by the total number of quality claims. Some are checklist or binary measures, such as whether all required execution settings are reported. Some are numerical error measures, such as the absolute difference between an observed value and a reference value. Some are distributional measures, such as the distance between an observed output distribution and a reference distribution. Some are statistical measures, such as confidence interval width or between-run variation. Others are robustness or sensitivity measures, such as the change in an output quantity under backend variation, shot-budget variation, or calibration drift.

The term \emph{candidate metric} therefore reflects two points. First, the metric is concrete enough to be computed, checked, or collected once its required context is provided. Second, the metric is not proposed as a universal standardized measure whose threshold and aggregation rule are fixed for all quantum software systems. Thresholds, weights, aggregation rules, and acceptance bounds should be defined in a tailored quality profile according to the application domain, quality priorities, and available evidence. This is consistent with the purpose of the proposed model as a base model for quality assessment rather than a fixed standard.

The representative candidate metrics in Table~\ref{table:criteria-candidate-metrics} are therefore organized according to these elements. For each criterion, the table identifies a candidate metric, the measurement object, the measurement function or counting rule, the required context and interpretation, and whether a reference or acceptance condition is normally needed. This structure is intended to make the metric layer explicit and usable while allowing project-specific instantiation.

\subsection{Types of Metrics in the Quantum Software Context}
\label{subsec:metric-types}

The candidate metrics used for quantum software quality do not have a single form. Although ratio and checklist metrics remain useful, they are not sufficient for capturing the quality concerns introduced by quantum computation. Quantum software quality assessment must also consider probabilistic behavior, statistical uncertainty, backend dependence, noise, transpilation effects, error mitigation, and hybrid quantum-classical optimization. Therefore, the metric layer of the proposed model includes several types of metrics. These types are not mutually exclusive; a concrete metric may combine more than one type.

\begin{itemize}
[leftmargin=15pt,itemsep=2.0pt,topsep=4pt,parsep=0pt]

    \item \textit{Structural and artifact metrics.}
    These metrics measure static properties of software artifacts. They can be collected from requirements, specifications, source programs, quantum circuits, transpiled circuits, documentation, test artifacts, and reproducibility packages. Examples include the number of unresolved assumptions, trace links between requirements and implementation artifacts, circuit depth, two-qubit gate count, number of backend-specific operations, number of unused code fragments, number of undocumented parameters, and number of missing execution settings. These metrics are useful because many quality problems can be detected before execution, especially those related to maintainability, portability, reproducibility, documentation, and resource use.

    \item \textit{Coverage and completeness metrics.}
    These metrics measure the extent to which required artifacts, assumptions, functions, tests, or quality evidence are present. They are often expressed as ratios, such as the number of traced requirements divided by the total number of requirements, the number of documented execution settings divided by the total number of required settings, or the number of tested circuit configurations divided by the total number of selected configurations. These metrics follow the classical tradition of ratio-based quality metrics, but their measurement objects must be extended to include quantum-specific artifacts such as circuit variants, measurement bases, backend configurations, shot budgets, mitigation settings, and calibration records.

    \item \textit{Checklist and binary metrics.}
    These metrics evaluate whether required properties, practices, or evidence items are present. A binary metric records whether a property holds or does not hold, while a checklist metric aggregates a set of binary or ordinal checks. For example, a reproducibility checklist may check whether the backend, shot count, random seeds, transpilation settings, mitigation configuration, software versions, and input problem instances are reported. A no-cloning compliance checklist may check whether a design, transformation, or distributed workflow avoids any requirement or assumption that arbitrary unknown quantum states can be perfectly duplicated. Checklist metrics are useful for criteria that are difficult to capture by a single numerical formula but can still be evaluated systematically.

    \item \textit{Numerical error metrics.}
    These metrics measure the deviation between an observed value and a reference value or acceptance bound. They are common in quantum algorithms and quantum simulation, where the expected result may be a reference energy, probability, expectation value, fidelity, or known analytical value. Typical examples include absolute value error, relative value error, energy error, expectation-value error, and deviation from a specified acceptance interval. Such metrics are reference-based and must state the reference value, how the observed value is estimated, and the execution setting under which the value is obtained.

\item \textit{Distributional metrics.}
These metrics characterize empirical output distributions or compare them with stated reference distributions or expected support properties. They are important because many quantum programs produce distributions over measurement results rather than single deterministic outputs. Examples include total variation distance, Hellinger distance, Kullback--Leibler divergence when well-defined, cross-entropy, heavy-output probability, and agreement with an expected support set. Their interpretation depends on the measurement basis, shot count, sampling uncertainty, and any reference distribution or expected support property. They should therefore be reported together with appropriate statistical uncertainty information.

    \item \textit{Statistical uncertainty and repeatability metrics.}
    These metrics measure the stability of observed results under repeated executions. Examples include confidence interval width, standard error, between-run variation, coefficient of variation, and pass-rate variation across repeated runs with nominally fixed settings.These metrics are central to quantum software because observed behavior depends on finite-shot sampling and may also be affected by device noise, calibration drift, and nondeterministic transpilation or optimization choices. Execution delays can matter indirectly when repeated runs occur under different time-dependent device conditions. A result that appears correct in one execution may be unreliable if it has large uncertainty or poor repeatability.

   \item \textit{Resource metrics.}
These metrics measure the computational and physical resources required by a quantum software artifact. Examples include logical qubit count, physical qubit requirement when applicable, circuit depth, two-qubit gate count, non-native gate count, number of measurements, shot count, number of circuit evaluations, optimizer iterations, memory use, and classical compute consumption. End-to-end runtime and queueing time are treated separately as execution-time measures rather than resource-utilization measures. For near-term quantum software, resource demands are directly connected to feasibility because qubit availability, circuit depth and associated noise exposure, and sampling overhead constrain what can be executed within stated bounds.

    \item \textit{Robustness and sensitivity metrics.}
    These metrics measure how much a quality-relevant result changes when the evaluation context changes. Examples include output variation under different shot budgets, transpilation seeds, optimization levels, backend choices, noise models, mitigation settings, calibration windows, or small perturbations of input parameters. These metrics are especially important for criteria such as calibration robustness, mitigation sensitivity, entanglement backend robustness, entanglement compilation robustness, backend portability, and constraint tolerance. They help distinguish software whose quality depends on a narrow and fragile setting from software whose behavior remains acceptable across reasonable execution conditions.

   \item \textit{Agreement and portability metrics.}
These metrics measure agreement among results or preservation of stated acceptance conditions across platforms, backends, simulators, or toolchains. Examples include cross-backend agreement rate, simulator--hardware deviation, transpiler-version agreement, and the proportion of target backends on which the program satisfies a stated acceptance condition. Such metrics are useful when evaluating backend portability and version compatibility and when assessing the factor-level concern of quantum hardware adaptability. They should state the compared backends or toolchains, the agreement or acceptance condition, and the allowable tolerance.

    \item \textit{Lifecycle evidence metrics.}
    These metrics measure whether quality evidence is maintained across development, execution, and evolution. Examples include the proportion of quality claims linked to supporting artifacts, the proportion of changed circuits for which affected metrics are re-evaluated, the number of stale execution records after backend recalibration, and the proportion of reproduced results that satisfy the original acceptance condition. These metrics are important because quantum software quality is not a one-time property of source code alone. It may need to be re-established when circuits are transformed, backends change, calibration data are updated, mitigation procedures are replaced, or optimizer and transpiler versions evolve.
\end{itemize}

These metric types provide a broader measurement basis than a metric set limited to classical ratio-based measures. This broader basis does not abandon the factor-criterion-metric structure; rather, it adapts the metric layer to the objects and behaviors that determine quantum software quality. These metric types provide the basis for the candidate metrics summarized in Table~\ref{table:criteria-candidate-metrics}.

\subsection{Representative Candidate Metrics}
\label{subsec:representative-metrics}

Table~\ref{table:criteria-candidate-metrics} presents representative candidate metrics for the quality criteria defined in Table~\ref{table:criteria-definitions}. The purpose of the table is to make the metric layer of the proposed model explicit. Each entry identifies a measurable or checkable way to evaluate a criterion, rather than merely naming a general measurement direction. The metrics are representative because different projects may instantiate additional metrics or refine the listed metrics according to their quality profile, application domain, available artifacts, and execution environment.

The table should be read together with the metric requirements described above. For each criterion, the table gives a candidate metric, the primary measurement object, a measurement function or counting rule, the required context for meaningful interpretation, and whether a reference value, reference distribution, oracle, expected property, or acceptance condition is normally required. This organization reflects the fact that a metric is useful only when its measurement object, measurement rule, and interpretation context are made explicit.

Several criteria may be supported by more than one metric. For example, correctness-related criteria may require both value-based and behavior-based evidence, while reproducibility-related criteria may require both artifact completeness and repeated-execution agreement. Conversely, the same metric may provide evidence for more than one criterion, but its interpretation may differ depending on the quality concern being assessed. For example, circuit depth may be interpreted as a resource-related metric, a feasibility indicator for constraint tolerance, or a risk indicator for noise-related reliability.

The table does not define fixed thresholds, weights, or factor-level scores. Such decisions are profile-dependent and should be specified when the base model is tailored to a concrete project. The base model provides a structured set of factors, criteria, and representative candidate metrics, while leaving project-specific thresholds, acceptance conditions, and aggregation rules to the quality profile.

\subsection{Metric Interpretation and Factor-Level Use}
\label{subsec:metric-interpretation}

A metric value is not a quality factor by itself. In the proposed model, metrics are used to evaluate criteria, and evaluations of those criteria provide evidence for assessing factors. Therefore, a metric should be interpreted through the criterion it supports and the quality profile in which it is used. For example, circuit depth may provide evidence for resource utilization, constraint tolerance, and noise-related reliability, but its meaning differs in each case. As a resource-utilization metric, smaller depth may indicate lower resource demand. As a constraint-tolerance metric, the same value may indicate whether the circuit can satisfy target backend constraints. As a reliability-related indicator, it may suggest exposure to decoherence or gate noise. The metric value is the same, but the quality interpretation depends on the criterion and factor under consideration.

This distinction is important because the relationship between metrics, criteria, and factors is many-to-many. A single criterion may require several metrics to capture different aspects of the same property. For instance, statistical repeatability may be evaluated by between-run variation, confidence interval width, and pass-rate variation across repeated executions. Conversely, a single metric may support several criteria when interpreted under different quality concerns. For this reason, metrics should not be treated as isolated numerical scores. They should be linked to the criteria they evaluate, the artifacts or executions from which they are collected, and the factor-level concerns they inform.

Some metrics are directly interpretable because their direction is clear. For example, smaller value error, smaller distributional distance, lower circuit depth, lower two-qubit gate count, and narrower confidence intervals usually indicate better quality with respect to the corresponding criterion. Other metrics require a stated threshold or acceptance condition. For example, a success probability may be acceptable only if it exceeds a domain-specific bound, and a cross-backend agreement rate may be meaningful only after defining the backends compared and the tolerance used to judge agreement. Checklist and coverage metrics also require explicit interpretation rules, because a high checklist score may still hide a missing item that is essential for a particular application.

Reference-based metrics require special care. Metrics such as value error, expectation-value error, fidelity loss, distributional distance, and oracle agreement depend on a reference value, reference distribution, expected property, or acceptance condition. The reference may come from an analytical result, a trusted simulator, a smaller exactly solvable instance, a classical baseline, a domain-specific physical model, or an independently validated implementation. The choice of reference should be recorded because different references may lead to different quality judgments. When no exact reference is available, the quality profile should specify how approximate references, consistency checks, metamorphic relations, or cross-execution agreement are used.

\clearpage
\onecolumn

\begingroup
\scriptsize

\setlength{\tabcolsep}{2pt}
\renewcommand{\arraystretch}{1.83}
\setlength{\LTleft}{-0.5cm}
\setlength{\LTright}{0pt}

\begin{longtable}{P{0.08\linewidth}P{0.12\linewidth}P{0.17\linewidth}P{0.36\linewidth}P{0.23\linewidth}P{0.05\linewidth}}
\caption{Representative Candidate Metrics for Quality Criteria in the Quantum Context}
\label{table:criteria-candidate-metrics}\\

\multicolumn{6}{p{1.0\linewidth}}{\textit{Note.} 
The column ``Ref.'' indicates whether the metric normally requires a reference value, reference distribution, oracle, expected property, or acceptance condition. ``Yes'' means that such a reference is normally required; ``No'' means that the metric can normally be computed without one; ``Profile'' means that the need for a reference or acceptance condition depends on the tailored quality profile.}\\[2mm]

\toprule
\textbf{Criterion}
& \textbf{Candidate Metric}
& \textbf{Measurement Object}
& \textbf{Measurement Function / Counting Rule}
& \textbf{Required Context and Interpretation}
& \textbf{Ref.} \\
\midrule
\endfirsthead

\caption[]{Representative Candidate Metrics for Quality Criteria in the Quantum Context (continued)}\\
\toprule
\textbf{Criterion}
& \textbf{Candidate Metric}
& \textbf{Measurement Object}
& \textbf{Measurement Function / Counting Rule}
& \textbf{Required Context and Interpretation}
& \textbf{Ref.} \\
\midrule
\endhead

\midrule
\multicolumn{6}{r}{Continued on next page}\\
\endfoot

\bottomrule
\endlastfoot

\multicolumn{6}{l}{\textit{Classical
(McCall) Criteria}}\\
\midrule

Traceability
& Quality-trace coverage
& Requirements, design artifacts, source code, circuits, tests, execution records
& $\frac{\#\text{quality-relevant claims linked to required artifacts and evidence}}{\#\text{quality-relevant claims}}$
& Larger values indicate stronger traceability. The trace scope should state which artifacts and evidence records are required.
& No \\

Completeness
& Required-evidence completeness
& Requirements, implementation artifacts, tests, documentation, execution records
& $\frac{\#\text{required functions, assumptions, settings, or evidence items present}}{\#\text{required items}}$
& Larger values indicate more complete implementation or evidence. Missing critical items should be reported separately even when the ratio is high.
& Profile \\

Consistency
& Cross-artifact consistency rate
& Requirements, design descriptions, code, circuits, configuration files, result reports
& $\frac{\#\text{checked artifact relations without inconsistency}}{\#\text{checked artifact relations}}$
& Larger values indicate stronger consistency. The checked relations may include names, parameter meanings, circuit versions, backend settings, and reported results.
& Profile \\

Accuracy
& Value error
& Observed numerical result, expectation value, energy estimate, probability estimate
& $|v_{\mathrm{obs}}-v_{\mathrm{ref}}|$ or $\frac{|v_{\mathrm{obs}}-v_{\mathrm{ref}}|}{|v_{\mathrm{ref}}|}$ when relative error is meaningful, where $v_{\mathrm{obs}}$ is the measured or estimated value and $v_{\mathrm{ref}}$ is the reference value
& Smaller values indicate higher accuracy. The reference value, estimator, shot count, backend, noise model, and mitigation setting should be recorded.
& Yes \\

Error tolerance
& Failure-containment pass rate
& Test executions under invalid inputs, noise perturbations, backend failures, or nonnominal conditions
& $\frac{\#\text{nonnominal executions satisfying the containment condition}}{\#\text{nonnominal executions}}$
& Larger values indicate better tolerance. The injected condition and expected safe behavior must be specified.
& \cyan{Yes} \\

Simplicity
& Normalized structural complexity score
& Source code, quantum circuits, hybrid workflow, configuration structure
& $C_{\mathrm{struct}}=\sum_i w_i \tilde{c}_i$, where $\tilde{c}_i$ is a selected complexity indicator normalized to a common profile-defined scale, $w_i \geq 0$, and $\sum_i w_i=1$. Indicators may include nesting depth, circuit depth, branch count, parameter count, or inter-component dependency count
& Smaller values indicate simpler structure. The selected indicators, normalization rules, and weights should be defined in the quality profile.
& Profile \\

Modularity
& Dependency-locality ratio
& Modules, quantum subroutines, circuit blocks, kernels, and hybrid components
& $\frac{\#\text{dependencies whose source and target lie within the same module}}{\#\text{all analyzed dependencies}}$
& Larger values indicate that dependencies are more strongly localized within module boundaries. Module boundaries and dependency types must be defined.
& No \\

Generality
& Evaluation-set applicability rate
& Algorithms, circuit templates, subroutines, parameterized kernels, APIs, and a declared finite evaluation set of target problem instances or configurations
& $\displaystyle
\frac{N_{\mathrm{supported}}}{N_{\mathrm{eval}}}$,
where $N_{\mathrm{supported}}$ is the number of elements in the declared evaluation set that are supported without rewriting core logic, and $N_{\mathrm{eval}}$ is the total number of elements in that evaluation set
& Larger values indicate broader applicability within the evaluated scope. The evaluation set should represent the intended problem family and relevant dimensions of variation, such as input instances, parameter settings, or deployment contexts.
& Profile \\

Expandability
& Extension-effort estimate
& Design, architecture, APIs, circuit templates, configuration files
& Number of modified components or changed interfaces required to add a stated function, backend, problem size, or measurement option
& Smaller values indicate easier expansion. The target extension scenario must be specified.
& Profile \\

Instrumentation
& Observability coverage
& Logs, counters, execution records, circuit statistics, optimizer traces, measurement records
& $\frac{\#\text{required observability items recorded}}{\#\text{required observability items}}$
& Larger values indicate stronger instrumentation. Required items may include raw counts, random seeds, parameters, backend, shot count, transpilation settings, and error messages.
& No \\

Self-descriptiveness
& Explanation completeness score
& Code comments, circuit annotations, API documentation, notebooks, experiment records
& Checklist score over required explanations, such as purpose, assumptions, inputs, outputs, parameters, qubit roles, measurement basis, and backend requirements
& Larger values indicate stronger self-descriptiveness. Critical missing explanations should be reported separately.
& No \\

Execution efficiency
& End-to-end processing time
& Program execution, circuit execution, backend jobs, and hybrid optimization workflow
& Elapsed time from workflow invocation to availability of a result satisfying the stated completion condition, with compilation or transpilation, backend execution, classical orchestration, post-processing, and queueing time reported separately when included
& Smaller values indicate better execution efficiency. The workload, simulator or backend, shot count, stopping condition, included processing stages, and treatment of queueing time must be stated.
& Profile \\

Storage efficiency
& Storage-cost measure
& Program artifacts, intermediate states, logs, result data, simulation data
& Memory use, result-storage size, or number of stored intermediate objects per run or per problem instance
& Smaller values indicate better storage efficiency. The artifact retention policy and execution scale should be stated.
& No \\

Access control
& Authorization-enforcement pass rate
& Authorization tests for repositories, execution services, credentials, backend accounts, result storage, and other protected resources
& $\frac{\#\text{authorization tests producing the specified allow-or-deny outcome}}{\#\text{authorization tests}}$
& Larger values indicate stronger access control. Each test should state the user or role, resource, requested operation, applicable policy, and expected authorization decision.
& \cyan{Yes} \\

Access audit
& Audit-record completeness
& Access logs, backend job records, repository logs, execution and result-storage logs
& $\frac{\#\text{security-relevant actions with complete audit records}}{\#\text{security-relevant actions}}$
& Larger values indicate stronger auditability. Required fields may include actor, time, resource, action, and outcome.
& No \\

Operability
& Task-success rate
& User tasks, execution workflows, job submission, result interpretation, error handling
& $\frac{\#\text{representative user tasks completed correctly}}{\#\text{representative user tasks}}$
& Larger values indicate better operability. User role, task set, environment, and allowed assistance should be specified.
& \cyan{Yes} \\

Training
& Training-task completion rate
& Tutorials, examples, onboarding tasks, teaching materials, user guides
& $\frac{\#\text{training tasks completed by target users}}{\#\text{training tasks}}$
& Larger values indicate better training support. The target user group and training tasks should be specified.
& Profile \\

Communica\-tiveness
& Message interpretability score
& Error messages, warnings, result reports, visualizations, logs, documentation
& Checklist or user-rating score over clarity, relevance, actionability, and correct use of quantum terminology
& Larger values indicate more useful communication. The evaluation should specify user role and message type.
& Profile \\

Software system independence
& Software-environment-specific dependency ratio
& Source code, build and environment files, runtime-service calls, library and toolchain dependencies, and execution scripts
& $\frac{\#\text{analyzed dependencies specific to a particular software environment}}{\#\text{analyzed software-environment dependencies}}$
& Smaller values indicate greater software system independence. The target software environments, analyzed dependency types, and environment-specific dependency classification rule should be stated.
& No \\

Machine independence
& Classical-host-specific dependency ratio
& Source code, build scripts, runtime configurations, native extensions, hardware-interface calls, and architecture-specific code
& $\displaystyle
\frac{N_{\mathrm{host}}}{N_{\mathrm{dep}}}$,
where $N_{\mathrm{host}}$ is the number of analyzed dependencies specific to a particular classical host hardware configuration, and $N_{\mathrm{dep}}$ is the total number of analyzed classical-host hardware dependencies
& Smaller values indicate greater machine independence. The target classical host hardware classes, analyzed dependency types, and classification rule should be stated.
& No \\

Communications commonality
& Standard-protocol usage rate
& Communication interfaces, transport protocols, job-submission mechanisms, callbacks, and result-retrieval mechanisms
& $\frac{\#\text{communication interfaces using standard or agreed protocols}}{\#\text{communication interfaces}}$
& Larger values indicate stronger communications commonality. The accepted protocols and interface mechanisms should be specified.
& No \\

Data commonality
& Data-format commonality rate
& Input data, result data, circuit representations, configuration files, and metadata
& $\frac{\#\text{data artifacts using standard or agreed representations}}{\#\text{data artifacts}}$
& Larger values indicate stronger data commonality. The accepted representations, formats, schemas, and required metadata fields should be stated.
& No \\

Conciseness
& Artifact redundancy rate
& Source code, circuit descriptions, configuration files, notebooks, and documentation
& $\frac{\#\text{duplicate, unused, or semantically redundant analyzed elements}}{\#\text{analyzed elements}}$
& Smaller values indicate better conciseness. The analyzed element type and redundancy-classification rule should be stated, and the metric should not reward obscure or under-documented artifacts.
& No \\

\midrule
\multicolumn{6}{l}{\textit{Classical
(Ext.) Criteria}}\\
\midrule

Resource utilization
& Resource-envelope satisfaction rate
& Program executions, circuit executions, backend jobs, simulation runs, and classical post-processing records
& $\frac{\#\text{executions satisfying the stated quantum and classical resource envelope}}{\#\text{evaluated executions}}$
& Larger values indicate better resource utilization with respect to stated requirements. The resource envelope may include logical or physical qubit requirements, circuit depth, two-qubit gate count, shot count, number of circuit evaluations, memory use, and classical compute consumption.
& \cyan{Yes} \\

Readability
& Readability review score
& Source code, circuit descriptions, configuration files, notebooks, comments, and documentation fragments
& Checklist or rating score over naming clarity, structural clarity, explanation of quantum assumptions, circuit readability, and ease of following the hybrid workflow
& Larger values indicate better readability. The reviewer role, artifact scope, checklist items, and scoring rule should be specified.
& Profile \\

Change locality
& Change-impact locality ratio
& Source modules, circuit components, configuration files, backend adapters, and orchestration logic
& $1-\frac{\#\text{artifacts changed outside the intended change scope}}{\#\text{artifacts affected by the change}}$
& Larger values indicate stronger change locality. The intended change scenario and affected artifact boundary must be stated.
& Profile \\

Documentation adequacy
& Required-documentation adequacy score
& User guides, developer notes, API documentation, experiment records, reproduction instructions, and result reports
& $\frac{\#\text{required documentation items judged adequate}}{\#\text{required documentation items}}$
& Larger values indicate more adequate documentation. Required items should cover purpose, assumptions, inputs, outputs, execution settings, limitations, evidence rules, and interpretation guidance.
& No \\

Tool support
& Quality-activity tool support rate
& Testing, debugging, analysis, execution, monitoring, visualization, packaging, and maintenance workflows
& $\frac{\#\text{selected quality activities with effective tool support}}{\#\text{selected quality activities}}$
& Larger values indicate stronger tool support. The selected activities, tool maturity, automation level, and integration points should be specified.
& No \\

Configurability
& Controlled-configuration coverage
& Configuration files, command-line options, runtime parameters, backend settings, mitigation settings, and optimizer settings
& $\frac{\#\text{quality-relevant settings exposed through controlled configuration}}{\#\text{quality-relevant settings}}$
& Larger values indicate better configurability. Settings that affect results but remain hidden, undocumented, or changed through ad hoc code edits should be treated as weak or missing.
& No \\

Reproducibility
& Reproduction agreement rate
& Repeated reproduction attempts, scripts, environment files, execution logs, raw counts, processed results, and reported claims
& $\frac{\#\text{reproduction attempts satisfying the stated agreement condition}}{\#\text{reproduction attempts}}$
& Larger values indicate stronger reproducibility. The dependency versions, backend or simulator, shot count, random seeds, transpilation settings, input instances, and agreement condition must be recorded.
& Yes \\

Recoverability
& Valid-state recovery rate
& Failed executions, interrupted backend jobs, invalid configurations, corrupted intermediate results, and restarted workflows
& $\frac{\#\text{failure scenarios recovered to a valid state}}{\#\text{tested failure scenarios}}$
& Larger values indicate better recoverability. The valid recovery state, recovery policy, failure scenarios, and maximum allowed recovery effort should be defined.
& \cyan{Yes} \\

Confidentiality
& Sensitive-asset protection coverage
& Credentials, access tokens, circuits, configurations, datasets, intermediate data, results, logs, and backend account information
& $\frac{\#\text{sensitive assets protected against unauthorized disclosure}}{\#\text{sensitive assets requiring protection}}$
& Larger values indicate stronger confidentiality. The sensitive asset set, threat boundary, storage locations, transmission channels, and protection mechanisms should be stated.
& Profile \\

Authenticity
& Artifact-origin verification rate
& Source code, circuit files, datasets, configuration files, result files, packages, containers, and execution records
& $\frac{\#\text{artifacts whose claimed identity and origin are verified}}{\#\text{artifacts requiring authenticity evidence}}$
& Larger values indicate stronger authenticity. Evidence may include digital signatures, authenticated provenance records, trusted repositories, or controlled release records. Unauthenticated checksums alone do not verify origin and should not be treated as authenticity evidence.
& No \\

Vulnerability resistance
& Security-test resistance rate
& Security test cases and attack or misuse scenarios derived from the stated threat model for source code, dependencies, APIs, execution services, backend access paths, containers, and hybrid workflow interfaces
& $\displaystyle
\frac{N_{\mathrm{resist}}}{N_{\mathrm{sec}}}$,
where $N_{\mathrm{resist}}$ is the number of executed security test cases or scenarios for which the specified security property remains satisfied, and $N_{\mathrm{sec}}$ is the total number of executed security test cases or scenarios
& Larger values indicate stronger resistance within the assessed threat model and test scope. The threat model, test or scenario set, expected secure behavior, assessment environment, and coverage should be stated. Passing the assessed scenarios does not imply the absence of vulnerabilities outside the evaluated scope.
& Yes \\

Hazard control
& Hazard mitigation coverage
& Requirements, design artifacts, execution constraints, monitoring rules, fallback procedures, and safety or security analyses
& $\frac{\#\text{identified hazards with defined controls or mitigations}}{\#\text{identified hazards}}$
& Larger values indicate stronger hazard control. The hazard identification method, severity classification, and required control evidence should be stated.
& Profile \\

Fail-safe behavior
& Safe-fallback pass rate
& Invalid inputs, backend failures, timeout events, failed mitigation, failed post-processing, and orchestration failures
& $\frac{N_{\mathrm{safe}}}{N_{\mathrm{failure}}}$, where $N_{\mathrm{safe}}$ is the number of failure scenarios leading to controlled termination, safe fallback, or a specified safe state, and $N_{\mathrm{failure}}$ is the total number of tested failure scenarios
& Larger values indicate stronger fail-safe behavior. The safe state, fallback policy, failure triggers, and conditions for avoiding misleading outputs should be specified.
& \cyan{Yes} \\

Interface conformance
& Interface-contract pass rate
& APIs, data formats, parameter passing, job submission, result retrieval, metadata exchange, and hybrid workflow interfaces
& $\frac{\#\text{interface tests satisfying the specified contract}}{\#\text{interface tests}}$
& Larger values indicate stronger interface conformance. The interface contract should define syntax, data format, units, parameter semantics, error handling, and result interpretation.
& Yes \\

Version compatibility
& Version-matrix pass rate
& Toolchain versions, quantum SDK versions, simulator versions, transpiler versions, backend API versions, and dependency sets
& $\frac{\#\text{tested version combinations satisfying the compatibility condition}}{\#\text{tested version combinations}}$
& Larger values indicate stronger version compatibility. The version matrix, dependency constraints, compatibility condition, and acceptable deviations should be recorded.
& \cyan{Yes} \\

Result transparency
& Result-interpretation completeness
& Result tables, plots, logs, notebooks, reports, raw counts, processed estimates, uncertainty records, and post-processing scripts
& $\frac{\#\text{reported results accompanied by required interpretation context}}{\#\text{reported results}}$
& Larger values indicate stronger result transparency. Required context should include backend or simulator, shot count, random-seed policy, transpilation settings, mitigation choices, uncertainty, aggregation, post-processing steps, and acceptance condition.
& No \\

Co-existence
& Shared-environment co-existence pass rate
& Executions in shared environments with other products or components and shared-resource usage records
& $\displaystyle
\frac{N_{\mathrm{pass}}}{N_{\mathrm{shared}}}$,
where $N_{\mathrm{pass}}$ is the number of evaluated shared-environment
executions satisfying both the required-function and non-interference
conditions, and $N_{\mathrm{shared}}$ is the total number of evaluated
shared-environment executions
& Larger values indicate stronger co-existence. The shared environment, co-running products or components, resource-sharing scenario, required-function condition, and non-interference condition should be stated.
& Yes \\

\midrule
\multicolumn{6}{l}{\textit{Quantum Criteria}}\\
\midrule

Entanglement stability
& Entanglement degradation measure
& Intended entangled states or subcircuits, validated entanglement witnesses or measures, and repeated execution records
& Change in a validated entanglement witness statistic or entanglement measure, or in an explicitly justified entanglement-certifying relation, under specified noise and decoherence conditions
& Smaller degradation or higher preservation rate indicates stronger entanglement stability. The required entanglement property, evidence measure or relation, qubit subset, reference or baseline, noise condition, shot count, and uncertainty procedure should be stated.
& Yes \\

Entanglement backend robustness
& Cross-backend entanglement variation
& Entanglement evidence collected for semantically equivalent executions across selected simulators, noisy simulators, or hardware backends
& Variation in a validated entanglement witness statistic or entanglement measure, or in an explicitly justified entanglement-certifying relation, across the selected backends
& Smaller variation indicates stronger preservation of the required entanglement property across backends. The entanglement property, logical-qubit subset, backend set, transpilation policy, shot count, uncertainty procedure, and acceptance tolerance should be stated.
& Profile \\

Entanglement compilation robustness
& Compilation entanglement-preservation gap
& Logical circuit and compiled or transpiled variants, logical-qubit mappings, and entanglement evidence at designated semantic checkpoints
& Difference in a validated entanglement witness statistic or entanglement measure, or in an explicitly justified entanglement-certifying relation, between semantically corresponding checkpoints before and after compilation
& Smaller gaps indicate stronger preservation of the required entanglement property through compilation. The entanglement property, comparison checkpoints, logical-qubit correspondence, compiler or transpiler version, optimization settings, backend target, and acceptance tolerance should be stated. Arbitrary gate-by-gate alignment between the logical and compiled circuits is not assumed.
& Yes \\

Backend portability
& Cross-backend acceptance rate
& Executions or simulations over selected target backends
& $\frac{\#\text{target backends satisfying the stated behavioral acceptance condition}}{\#\text{target backends}}$
& Larger values indicate stronger backend portability. The backend set, adaptation allowance, transpilation policy, shot count, noise assumptions, and acceptance condition must be specified.
& \cyan{Yes} \\

Calibration robustness
& Calibration-window stability
& Hardware executions, backend calibration records, and repeated metric values over time
& Change in a selected quality metric across calibration windows, such as value error, distributional distance, pass rate, or estimated expectation value
& Smaller change indicates stronger calibration robustness. The backend, calibration timestamps, execution windows, selected metric, and allowed drift should be recorded.
& Profile \\

Constraint tolerance
& Device-constraint satisfaction rate
& Logical circuits, transpiled circuits, backend constraints, coupling maps, basis gates, qubit limits, and execution constraints
& $\displaystyle
\frac{N_{\mathrm{sat}}}{N_{\mathrm{eval}}}$,
where $N_{\mathrm{sat}}$ is the number of evaluated constraint settings satisfying the acceptance condition, and $N_{\mathrm{eval}}$ is the total number of evaluated constraint settings
& Larger values indicate stronger constraint tolerance. The evaluated constraints may include connectivity, native gate set, available-qubit count, circuit-depth limit, and error budget.
& Yes \\

Orchestration correctness
& Orchestration-contract pass rate
& Hybrid workflow, classical control logic, quantum job submission, result aggregation, parameter updates, and decision logic
& $\frac{N_{\mathrm{pass}}}{N_{\mathrm{workflow}}}$, where $N_{\mathrm{pass}}$ is the number of workflow tests satisfying the intended control-flow, data-flow, and result-handling contract, and $N_{\mathrm{workflow}}$ is the total number of workflow tests
& Larger values indicate stronger orchestration correctness. The intended workflow contract, component invocation order, data dependencies, and result-handling rules must be specified.
& Yes \\

Synchronization correctness
& Synchronization-violation rate
& Quantum job scheduling, measurement-dependent actions, asynchronous callbacks, parameter updates, and classical post-processing steps
& $\frac{\#\text{observed ordering, timing, or dependency violations}}{\#\text{checked synchronization relations}}$
& Smaller values indicate stronger synchronization correctness. The required ordering, timing, dependency relations, timeout policy, and concurrency model should be stated.
& Yes \\

Orchestration robustness
& Workflow disruption tolerance rate
& Hybrid workflow executions under delays, retries, communication disruptions, backend queue changes, and transient failures
& $\frac{N_{\mathrm{robust}}}{N_{\mathrm{disrupted}}}$, where $N_{\mathrm{robust}}$ is the number of disrupted workflow executions that continue to satisfy stated acceptance conditions, recover within specified bounds, or enter a defined safe or controlled-degradation state, and $N_{\mathrm{disrupted}}$ is the total number of disrupted workflow executions
& Larger values indicate stronger orchestration robustness. The disruption model, retry policy, fallback behavior, and acceptable degradation condition must be specified.
& \cyan{Yes} \\

Measurement bias
& Measurement-bias magnitude
& Measurement calibration or reference circuits, measurement outcome distributions, calibration data, and repeated measurements
& Estimated systematic deviation attributable to the measurement process, obtained using a stated calibration or reference procedure after accounting for expected finite-shot variation
& Smaller values indicate less measurement bias. The calibration or reference procedure should be designed to isolate measurement-related effects and should state the measurement basis, shot count, estimator, and uncertainty procedure.
& Yes \\

Statistical repeatability
& Between-run statistical variation
& Repeated executions under nominally identical conditions
& Standard deviation, confidence-interval width, coefficient of variation, distributional variation, or pass-rate variation over repeated runs
& Smaller variation indicates stronger statistical repeatability. The repeated-run protocol, backend, shot count, random seeds, transpilation settings, mitigation settings, and time window must be stated.
& No \\

Mitigation sensitivity
& Maximum mitigation-induced deviation
& Results obtained under a designated baseline mitigation configuration and a specified set of alternative mitigation and post-processing configurations
& $\max_{m\in\mathcal{M}} d(r_m,r_{m_0})$, where $m_0$ is the designated baseline configuration, $r_m$ is the result obtained under configuration $m$, and $d$ is a stated distance or error measure appropriate to the result type
& Smaller values indicate lower mitigation sensitivity. The configuration set, baseline configuration, distance measure, fixed execution conditions, and uncertainty procedure should be stated.
& \cyan{Yes} \\

Unitarity preservation
& Unitary-region conformance rate
& Declared unitary regions and their representations before and after program or toolchain transformations
& $\displaystyle
\frac{N_{\mathrm{unitary}}}{N_{\mathrm{declared}}}$,
where $N_{\mathrm{unitary}}$ is the number of declared unitary regions whose effective transformations satisfy the stated unitarity check, and $N_{\mathrm{declared}}$ is the total number of declared unitary regions. When a matrix representation $U$ is available, conformance may be checked using $\|U^\dagger U-I\| \leq \epsilon$, with a stated matrix norm and tolerance $\epsilon$
& Larger values indicate stronger unitarity preservation. The declared unitary regions, transformation stage, semantic representation, checking method, and numerical tolerance when applicable should be stated.
& Yes \\

Measurement semantics soundness
& Measurement-semantics conformance rate
& Measurement basis, measurement placement, outcome labels, post-measurement state evolution, and measurement-conditioned classical control
& $\frac{\#\text{measurement-related checks conforming to the intended semantics}}{\#\text{measurement-related checks}}$
& Larger values indicate stronger measurement semantics soundness. The specified semantics should cover basis, placement, outcome interpretation, post-measurement state evolution, and measurement-conditioned classical control.
& Yes \\

No-cloning compliance
& No-cloning conformance violation count
& Quantum-data abstractions, interfaces, program transformations, and distributed quantum workflows
& Number of violations of explicitly defined no-cloning conformance rules, such as an abstraction, interface, or transformation requiring perfect duplication of an arbitrary unknown quantum state
& Smaller values indicate stronger no-cloning compliance. The conformance rules and analysis scope should be stated. Copying classical measurement results or independently preparing multiple systems in the same known state should not be counted as violations.
& Yes \\

Assumption preservation
& Declared-assumption preservation rate
& Requirements, algorithm design, implementation, compilation or transpilation results, execution settings, and result interpretation records
& $\frac{\#\text{declared assumptions preserved across transformation and execution stages}}{\#\text{declared assumptions}}$
& Larger values indicate stronger assumption preservation. The assumption set may include physical, semantic, backend, noise, topology, precision, measurement, and independence assumptions.
& \cyan{Yes} \\

\end{longtable}
\endgroup

\clearpage
\twocolumn


Quantum software metrics also depend strongly on execution context. A metric collected under one backend, shot count, transpilation setting, mitigation configuration, or calibration window may not be comparable with a metric collected under another setting. For example, two circuits with the same logical behavior may have different transpiled depth and two-qubit gate count on different devices. A result that appears accurate under an ideal simulator may become unstable under a noisy backend. A mitigation procedure may improve a value error while increasing variance or introducing sensitivity to calibration data. Therefore, metric interpretation should always include the evaluation context required by the metric definition.

The proposed model does not prescribe a universal aggregation formula from metrics to criteria or from criteria to factors. Such aggregation is necessarily profile-dependent. Different projects may assign different priorities to factors such as correctness, reliability, maintainability, and efficiency and to supporting criteria such as backend portability, constraint tolerance, and reproducibility. They may also use different thresholds, weights, confidence requirements, and reassessment triggers. For this reason, factor-level assessment should be performed through a tailored quality profile that specifies the selected factors, selected criteria, selected metrics, required evidence, acceptance conditions, and aggregation rules.

The model provides a structured set of quality factors and criteria, together with representative candidate metrics, but it does not impose a single scoring scheme for all quantum software. A project may instantiate the candidate metrics without changing their measurement rules, refine those rules, add domain-specific metrics, or exclude metrics that are irrelevant to its quality goals. What is required is that each selected metric be explicitly defined and interpreted in relation to the criterion it supports.

Table~\ref{table:criteria-candidate-metrics} follows this principle by identifying measurable or checkable evidence for the criteria in Table~\ref{table:criteria-definitions}, while leaving thresholds, weights, acceptance conditions, and aggregation rules to the tailored quality profile.

\section{Using the Proposed Model Across the Quantum Software Lifecycle}
\label{sec:lifecycle}

This section illustrates how the factor set in Table~\ref{table:qs-factors} and the set of quality criteria in Table~\ref{table:criteria-definitions} can be used across the five-phase quantum software lifecycle introduced in~\cite{zhao2020quantum}. We use the core factor-to-criterion mappings in Table~\ref{table:factor-criterion-mapping} and the representative candidate metrics in Section~\ref{sec:metrics} (Table~\ref{table:criteria-candidate-metrics}) as a common structure for organizing evidence. The discussion is descriptive rather than prescriptive, because lifecycle practices vary across projects, domains, and platforms. Our goal is to show how the proposed model can organize quality-related reasoning, evidence collection, and reassessment across lifecycle stages, rather than to define a new lifecycle process or methodology.

A central point is that quality assessment in quantum software should not be treated as an activity that starts only after implementation. Many quality concerns originate earlier, for example, in how requirements are stated, how assumptions are documented, how hybrid workflows are designed, and how configuration and provenance are managed. Conversely, evidence collected during testing and maintenance, including post-deployment operation, often feeds back into earlier lifecycle artifacts.

\subsection{Quality Concerns Across Lifecycle Stages}

Quality assessment begins with quantum software requirements analysis and continues through design, implementation, testing, and maintenance. For quality assessment purposes, the maintenance discussion below also includes post-deployment operation. Across these stages, the proposed model is not used as a fixed checklist. Rather, it helps identify which factors are being prioritized, which criteria make those factors assessable, and what kinds of evidence should be collected or preserved at each stage. The corresponding criteria and representative candidate metrics are defined in Sections~\ref{sec:criteria} and~\ref{sec:metrics} and Tables~\ref{table:factor-criterion-mapping} and~\ref{table:criteria-candidate-metrics}.

\vspace{2mm}
\noindent
\textbf{Requirements Analysis Phase.}
During requirements analysis, the main task is to state quality expectations in a form that can later be traced to design choices, implementation artifacts, execution settings, and test evidence. Factors such as \emph{correctness}, \emph{functionality}, \emph{reliability}, and \emph{documentation}, together with criteria such as \emph{traceability}, \emph{completeness}, \emph{reproducibility}, and \emph{result transparency}, provide the main assessment focus. For quantum software, requirements should specify expected functional behavior together with the assumptions under which results will be interpreted, such as backend class, allowable noise conditions, shot budget, mitigation policy, and acceptable numerical error, statistical uncertainty, or distributional deviation. When the software is part of a hybrid workflow, requirements should also clarify the roles of classical orchestration, data exchange, and measurement-conditioned control.

This stage should also define what later evidence will count as acceptable support for quality claims. For example, a requirement may specify an acceptable energy error, success probability, distributional distance, resource envelope, portability target across backends, or reporting obligation for uncertainty and execution settings. Quantum-specific concerns, including factor-level concerns such as quantum hardware adaptability and physical validity and criterion-level concerns such as entanglement stability, measurement bias, statistical repeatability, mitigation sensitivity, backend portability, constraint tolerance, and assumption preservation, should be recognized here when they are relevant to the intended quality claims, even if they are evaluated only during testing or operation. Making these concerns explicit early helps avoid requirements that appear meaningful in abstract terms but cannot be assessed under realistic execution conditions.

\vspace{2mm}
\noindent
\textbf{Design Phase.}
During design, the model helps determine whether the chosen structure will support later assessment or make it unnecessarily fragile. Factors such as \emph{maintainability}, \emph{flexibility}, \emph{interoperability}, and \emph{physical validity} are supported by criteria such as \emph{modularity}, \emph{simplicity}, \emph{self-descriptiveness}, \emph{interface conformance}, \emph{configurability}, and \emph{assumption preservation}. In quantum software, design quality is not limited to circuit decomposition. It also includes the separation between algorithm logic and execution support code, the design of classical orchestration logic, the representation of circuit and configuration artifacts, and the boundaries between quantum and classical responsibilities.

Several quantum-specific criteria should already be considered at this stage. Assumptions about measurement placement, ancilla allocation and required uncomputation or reset, unitary regions, qubit mapping, backend constraints, and no-cloning compliance may be invalidated later if the design does not make them explicit. For hybrid workflows, \emph{orchestration correctness} and \emph{synchronization correctness} are also design concerns, because ordering, data dependencies, and measurement-conditioned actions are often fixed before testing begins. Candidate evidence at this stage may include coupling and cohesion measures, interface definitions, change-isolation measures, configuration boundaries, design reviews of circuit abstractions, and early conformance checks for representations and workflow structure.

\vspace{2mm}
\noindent
\textbf{Implementation Phase.}
During implementation, the model becomes concrete through the artifacts that later support assessment, including source modules, circuit definitions, configuration files, compiled circuits, logs, and run metadata. Factors such as \emph{maintainability}, \emph{supportability}, \emph{manageability}, and \emph{usability} are supported by criteria such as \emph{consistency}, \emph{instrumentation}, \emph{self-descriptiveness}, \emph{tool support}, \emph{documentation adequacy}, \emph{reproducibility}, and \emph{result transparency}. In quantum software, implementation quality depends not only on whether code and circuits are written correctly, but also on whether execution settings, compiled artifacts, backend configurations, and acceptance conditions are recorded in a form that later assessment can use.

This stage is especially important for preserving evidence that would otherwise be lost. Run metadata, transpilation reports, random seeds, shot count, mitigation settings, compiled circuits, and configuration snapshots may be needed to diagnose failures, reproduce results, or explain differences between simulator and hardware executions. Documentation and training also begin to affect quality at this stage, because developers and users must be able to re-execute workflows, change controlled settings, and interpret probabilistic outputs without relying on implicit knowledge.

\vspace{2mm}
\noindent
\textbf{Testing Phase.}
During testing, selected candidate metrics are instantiated and applied under controlled execution conditions to evaluate the selected criteria. Factors such as \emph{reliability}, \emph{testability}, \emph{verifiability}, and \emph{supportability} become central, while criteria such as \emph{instrumentation}, \emph{traceability}, \emph{reproducibility}, \emph{statistical repeatability}, \emph{measurement bias}, \emph{mitigation sensitivity}, and \emph{result transparency} determine how test evidence should be produced and interpreted. Quantum software testing often involves repeated execution, explicit control of shot counts and random seeds, and, where relevant, simulator--device comparisons, together with careful treatment of statistical uncertainty.

Many quantum-specific criteria are evaluated most directly during testing, but their evaluation should remain linked to earlier assumptions. Candidate evidence may include measured variation in the entanglement properties required for the intended computation across repeated executions, estimates of measurement bias, between-run variation under nominally identical conditions, outcome variation across controlled mitigation choices, conformance results for hybrid orchestration logic, and differences between simulator and backend behavior. Testing often produces the richest body of quality evidence, but this evidence is also highly context-sensitive. It should therefore remain linked to backend identifiers, transpilation options, random seeds, shot budgets, mitigation settings, calibration windows, and the requirements that define acceptable results.

\vspace{2mm}
\noindent
\textbf{Maintenance Phase.}
For the purposes of quality assessment, this phase also includes post-deployment operation. Maintenance and evolution are especially important in quantum software because hardware platforms, compilers, toolchains, and service policies change rapidly. Software that behaves correctly in one environment may become brittle, inefficient, or misleading in another. Factors such as \emph{modifiability}, \emph{portability}, \emph{compatibility}, \emph{survivability}, \emph{manageability}, and \emph{quantum hardware adaptability} are therefore supported by criteria such as \emph{change locality}, \emph{version compatibility}, \emph{interface conformance}, \emph{documentation adequacy}, \emph{recoverability}, \emph{calibration robustness}, \emph{constraint tolerance}, and \emph{backend portability}.

Post-deployment operation introduces additional concerns that may not appear during isolated development. Cloud-based quantum execution requires controlled handling of access permissions, credentials, audit logs, queueing effects, calibration drift, and time-varying backend conditions. Criteria such as \emph{access control}, \emph{access audit}, \emph{operability}, \emph{confidentiality}, \emph{authenticity}, \emph{recoverability}, \emph{execution efficiency}, and \emph{result transparency} become important for deciding whether the software can be operated in a controlled and trustworthy way under realistic conditions. For hybrid workflows, operational quality also depends on whether service failures, delayed jobs, invalid results, and changing execution conditions are handled without producing misleading outputs.

Maintenance should also reassess whether earlier evidence remains valid. When a backend, transpiler, mitigation procedure, dependency version, calibration regime, or execution policy changes, previous claims about correctness, reliability, portability, or measurement-related quality may no longer hold. Candidate evidence at this stage may include access violation counts, audit-log completeness, provenance verification coverage, retry success rates under transient failures, reporting completeness for operational results, drift sensitivity across calibration periods, change-impact spread, regression rates after dependency upgrades, compatibility pass rates across toolchain versions, documentation update completeness, and successful re-execution after backend migration. From a lifecycle perspective, maintenance includes changing the software and preserving or re-establishing the validity of prior quality evidence under evolving conditions.

\subsection{Interactions Between Classical and Quantum-Specific Quality Concerns}

The lifecycle view also shows that classical and quantum-specific concerns should not be assessed independently. Classical criteria often determine whether quantum-specific evidence can be interpreted at all. For example, \emph{traceability} links evidence about measurement bias or mitigation sensitivity to the requirement, backend, shot budget, mitigation setting, and execution period under which it was obtained. \emph{Instrumentation} records random seeds, transpilation options, calibration windows, compiled circuits, and run metadata; without such records, evidence about statistical repeatability, calibration robustness, entanglement stability, or entanglement backend robustness is difficult to interpret. Similarly, \emph{documentation adequacy} and \emph{self-descriptiveness} make explicit the assumptions needed to understand probabilistic outputs, rather than merely making artifacts easier to read.

The reverse dependency is also important. Quantum-specific criteria can reveal limitations in software that appears well structured from a classical perspective. A workflow may be modular, readable, and configurable, yet still produce weak quality claims if backend-specific connectivity, calibration drift, or mitigation choices substantially change the observed distribution. A circuit transformation may preserve a clean module structure, yet still require reassessment if it changes the entanglement properties required for the intended computation, measurement semantics, or assumptions about unitary regions. For hybrid workflows, orchestration and synchronization concerns may also expose faults that are not visible from static code structure alone.

These interactions explain why the proposed model uses many-to-many mappings rather than a strict tree. A single criterion may support several factors, and evidence collected for one concern may strengthen, qualify, or weaken claims about another concern. The purpose of the model is therefore not only to list factors, criteria, and candidate metrics, but also to help users maintain the links among quality claims, quality criteria, evaluation context, and evolving execution conditions.

\subsection{Illustrative Quality Profile for a VQA Project}
\label{subsec:vqa-example}

Consider a variational quantum algorithm (VQA)~\cite{cerezo2021variational} project that estimates a molecular ground-state energy using a VQE-style workflow. The project maps a molecular Hamiltonian into a weighted sum of Pauli terms,
\begin{equation}
    \hat{H}=\sum_{j=1}^{M} c_j P_j,
\end{equation}
prepares trial states using a parameterized ansatz, estimates Pauli-term expectation values from repeated measurements, and uses a classical optimizer to update circuit parameters. The resulting energy estimate is typically computed as
\begin{equation}
    E(\theta)=\sum_{j=1}^{M} c_j \langle P_j\rangle_\theta,
\end{equation}
where $M$ is the number of Pauli terms, $c_j$ is the coefficient of the Pauli term $P_j$, $\langle P_j\rangle_\theta$ is the expectation value estimated from measurements of the ansatz state parameterized by $\theta$, and $\theta$ denotes the ansatz parameters. The context profile should specify the target molecule, molecular geometry, basis set, qubit mapping, reference or benchmark value used for assessment, acceptable energy error, Hamiltonian representation, ansatz family and depth, optimizer settings, initial parameters, stopping conditions, backend class, shot budget or shot-allocation policy, transpilation settings, measurement grouping strategy, mitigation policy, and the assumptions under which the final energy estimate is to be interpreted. These choices define the conditions under which quality claims about the VQA result are meaningful.

For such a project, a tailored quality profile may prioritize factors such as \emph{correctness}, \emph{reliability}, \emph{efficiency}, \emph{verifiability}, \emph{physical validity}, and \emph{quantum hardware adaptability}. The corresponding criteria may include \emph{traceability}, \emph{accuracy}, \emph{reproducibility}, \emph{statistical repeatability}, \emph{measurement bias}, \emph{mitigation sensitivity}, \emph{execution efficiency}, \emph{resource utilization}, \emph{backend portability}, and \emph{assumption preservation}. These criteria are selected because the quality of a VQA result depends not only on whether the circuit is implemented as intended, but also on whether the Hamiltonian terms are measured correctly, whether the optimizer reaches a stable solution, whether the estimated energy remains within an acceptable error bound, and whether backend constraints, transpilation choices, shot allocation, and mitigation settings preserve the assumptions used to interpret the result.

The selected criteria can then be evaluated using candidate metrics appropriate to this profile. For example, \emph{accuracy} may be evaluated by the deviation between the estimated energy and a stated reference value or acceptance bound. \emph{Statistical repeatability} may be evaluated by the variation of the final energy estimate across repeated runs under nominally identical execution conditions, with shot count, initial parameters, optimizer settings, backend conditions, and the random-seed policy explicitly controlled. \emph{Measurement bias} may be evaluated using calibration or reference circuits, or another stated measurement-bias estimation procedure, to estimate systematic measurement-related deviation in Pauli-term expectation values after accounting for expected finite-shot variation, while \emph{mitigation sensitivity} may be evaluated by the change in the final energy estimate or confidence interval across mitigation configurations. \emph{Backend portability} may be evaluated by the proportion of eligible backends on which the stated behavioral acceptance conditions are satisfied with no more than the allowed adaptation, while differences in estimated energy, uncertainty, circuit depth, gate counts, and shot use are reported as supporting context. \emph{Execution efficiency} may be evaluated using end-to-end processing time and its relevant components, while \emph{resource utilization} may be evaluated using circuit evaluations, shot count, qubit requirements, circuit depth, critical gate counts, memory use, and classical compute consumption. \emph{Assumption preservation} may be evaluated by checking whether assumptions about the ansatz structure, parameter binding, qubit mapping, Hamiltonian decomposition, measurement basis, backend capabilities, and mitigation procedure remain valid after transpilation and execution.

The evidence collected for this profile should be linked to the VQA artifacts and settings that give it meaning, including the Hamiltonian decomposition, ansatz definition, optimizer trace, measurement schedule, compiled circuits, backend configuration, mitigation settings, and run metadata. Testing may then focus on repeated executions, controlled changes in mitigation choices, optimizer stability under different initializations, and behavior across eligible backends.

If the project later changes its backend, transpiler version, ansatz, optimizer, initial-parameter strategy, measurement grouping, mitigation procedure, shot-allocation policy, qubit mapping, or Hamiltonian decomposition, earlier evidence cannot automatically be assumed to remain valid. The affected quality claims should be reassessed under the revised context profile using the relevant criteria and candidate metrics. This example illustrates how the proposed base quality model can be tailored to produce a project-specific quality profile for selecting evidence and determining when reassessment is required for a VQA-based molecular-energy estimation task.

\section{Discussion and Open Challenges}
\label{sec:discussion-challenge}

The proposed model is intended as a base quality model rather than a fixed standard. Several open challenges remain in applying such a model consistently across projects, platforms, and toolchains. The main challenges concern how to develop shared conventions for interpreting and reporting metrics, how to empirically validate the proposed factors, criteria, mappings, and candidate metrics, how to support assessment with practical tools, and how to maintain meaningful quality assessment under rapidly changing hardware and software environments.

\subsection{From Candidate Metrics to Common Measurement Conventions}

One challenge is to establish shared measurement conventions for the representative candidate metrics. A metric name or formula alone is insufficient for comparison unless the measured artifact or behavior, execution conditions, aggregation procedure, reference condition, uncertainty treatment, and reporting format are also specified. Otherwise, two studies may report similarly named metrics while measuring substantially different properties or operating under incomparable assumptions.

This issue is especially important for criteria such as \emph{measurement bias}, \emph{statistical repeatability}, \emph{mitigation sensitivity}, \emph{entanglement backend robustness}, and \emph{calibration robustness}. Similar issues also arise for extended classical criteria such as \emph{reproducibility} and \emph{result transparency}, whose evaluation depends on how execution evidence, assumptions, and reporting information are recorded. The resulting metric values may depend strongly on the backend, transpilation choices, shot count, random seeds, mitigation settings, and execution or calibration period. Future work should therefore develop conventions for recording the evaluation context, reporting uncertainty, defining suitable reference or acceptance conditions, and determining when the results obtained in different environments can be compared.

\subsection{Empirical Validation of the Model}

The proposed model also requires empirical validation. The lifecycle discussion and illustrative VQA profile presented in this paper show how the proposed base quality model may be applied, but they do not constitute validation. It remains necessary to examine whether the factor set is sufficiently broad while still practical, whether evaluations of the criteria provide useful evidence for the factors to which they are mapped, and whether the representative candidate metrics capture meaningful and interpretable aspects of software quality.

Such an evaluation should combine expert feedback, case studies involving representative quantum software systems, and empirical studies using real software artifacts and execution environments. Because quantum software behavior is probabilistic and environment-dependent, the evaluation should also consider different backends, workloads, execution periods, and toolchain versions. These studies can help determine which parts of the model are stable across contexts, whether the candidate metrics are sufficiently sensitive to meaningful quality differences, and which factors, criteria, mappings, or candidate metrics require further refinement.

\subsection{Tool Support for Quality Assessment}

Practical use of the model will also require integrated tool support for collecting, linking, and maintaining quality evidence. Existing quantum software tools mainly support individual activities such as compilation, execution, testing, debugging, or monitoring. Much less support exists for relating evidence from these activities to quality factors, criteria, requirements, and execution assumptions across lifecycle stages.

Such support would need to capture provenance for source programs, compiled circuits, configurations, backends, calibration periods, mitigation choices, execution logs, and results. It should also help users identify which quality claims are supported by the available evidence, which assumptions apply, and which environmental changes require reassessment. Support for configuration auditing and controlled re-execution would also help determine whether earlier results can be reproduced or reassessed under revised conditions. Integration with version control and continuous integration workflows would make it possible to treat quality evidence as an evolving project artifact rather than as isolated results produced by individual tools.

\subsection{Maintaining Quality Evidence Under Hardware and Toolchain Evolution}

Maintaining the validity of quality evidence as quantum hardware and software environments evolve presents another challenge. Evidence obtained under one backend, calibration period, compiler or transpiler version, native gate set, mitigation procedure, or platform policy may no longer support the same quality claim after those conditions change. Quality assessment must therefore record not only a measured result but also the environment and assumptions under which that result was obtained.

This issue affects factors such as \emph{portability}, \emph{survivability}, \emph{manageability}, and \emph{quantum hardware adaptability}, as well as criteria such as \emph{reproducibility}, \emph{version compatibility}, \emph{entanglement backend robustness}, and \emph{calibration robustness}. However, the problem is broader than any single factor or criterion: requirements, design assumptions, test evidence, and deployment conditions may evolve at different times and invalidate different parts of an earlier assessment.

A practical assessment process should therefore associate quality evidence with its provenance and applicable context, identify changes that trigger reassessment, and distinguish evidence that remains valid from evidence that must be reproduced or reinterpreted. In this sense, quality assessment for quantum software is not a one-time activity, but an ongoing process of maintaining and revising evidence as the execution environment changes.

\section{Related Work}
\label{sec:related-work}

The classical software quality models and standards that provide the foundation of our work have already been summarized in Section~\ref{subsec:classical-quality-models} and Table~\ref{table:classical-factor-models}. This section therefore focuses on work that addresses quality-related concerns specifically in quantum software.

At a broad level, previous studies have established quantum software engineering as a research area and identified software quality, testing, verification, maintenance, and lifecycle support as important concerns~\cite{murillo2025roadmap,zhao2020quantum,ali2022software,piattini2021toward}. These studies motivate treating quantum software quality as a software engineering problem, not merely as a matter of quantum hardware performance or reliability.

Requirements-oriented work provides more direct evidence that software quality concerns arise early in quantum software development. Saraiva \emph{et al.}~\cite{saraiva2021non} derive five generic non-functional requirements from constraints encountered when executing programs on real quantum devices. These requirements concern the available number of qubits, circuit depth, T-gate count, qubit connectivity, and the availability of native gates, and are related mainly to the performance-efficiency and reliability characteristics of ISO/IEC~25010. Yue \emph{et al.}~\cite{yue2023towards} examine how requirements engineering should be adapted to quantum software. They distinguish functional and extra-functional requirements as well as quantum, classical, and hybrid requirements, and discuss quality-related concerns including portability, performance, reliability, scalability, maintainability, and reusability. Sep\'ulveda \emph{et al.}~\cite{sepulveda2024systematic} provide a broader systematic review of requirements engineering for quantum computing. Their analysis identifies challenges such as defining quantum-specific and hybrid requirements, coping with rapid technology evolution, and addressing security, testing, and verification, and explicitly relates requirements-engineering challenges, advances, and future directions to ISO/IEC~25010 quality characteristics. These studies show that quality concerns are already embedded in quantum software requirements engineering, although they do not provide a general software quality model that organizes those concerns across factors, criteria, and metrics.

Work directly addressing quantum software quality is still relatively limited. Verdugo \emph{et al.}~\cite{verduro2021software} distinguish the quality of quantum hardware, quantum software platforms, and quantum software products. They argue that a quantum software quality environment should include at least a quality model, a set of software metrics, and supporting tools, and propose that existing software quality characteristics should be reconsidered and adapted to the quantum context while new quantum-specific characteristics may also be required. Their discussion highlights, among other issues, understandability, accuracy and precision under noisy execution, platform dependence, and quantum-specific concepts such as entanglement, superposition, and no-cloning. This work is close in motivation to ours, but its main contribution is a research agenda and the case for a quantum software quality environment; it does not construct a detailed base quality model.

Several studies have addressed the measurement side of quantum software quality. Zhao~\cite{zhao2021some} proposes size and structure metrics for quantum software at several abstraction levels, including source code, architectural and detailed design, specification, control-flow structure, and information flow. The work extends classical measures such as lines of code, Halstead measures, McCabe's cyclomatic complexity, and Henry--Kafura information-flow metrics to quantum software, while also introducing measures for quantum-specific program and design elements. Cruz-Lemus \emph{et al.}~\cite{cruz2021towards} focus more specifically on the understandability of quantum circuits and propose metrics based on circuit width and depth, circuit density, gate types, controlled operations, oracles, measurements, and ancilla qubits. Their initial proposal identifies understandability as an important quality concern related to maintenance and reuse, but leaves empirical validation for subsequent work. In later work, Cruz-Lemus \emph{et al.}~\cite{cruz2024quantum} further develop this metric set, report an empirical study of the relationship between the proposed metrics and circuit understandability, and present the QMetrics prototype for automated metric calculation and visualization. These studies show that software-oriented metrics can be defined for quantum artifacts. Their scope, however, is limited to particular internal attributes or specific quality concerns; they do not provide a general quality model relating a broader set of quality factors, criteria, and candidate metrics.

More recently, Yousuf and Sofi~\cite{yousuf2025characterizing} provide empirical evidence about the quality concerns that arise in real quantum software repositories. Their study analyzes 32,296 issues from 123 open-source projects across multiple categories of quantum software and examines bug types, severity, testing practices, documentation, code complexity, and affected quality attributes. Their results show that usability, maintainability, and interoperability are among the quality attributes frequently affected by defects, while quantum-specific bugs have a substantial impact on concerns such as performance, reliability, maintainability, and correctness-related behavior. They also report relationships between testing practices, documentation, complexity, and defect incidence. These results provide empirical evidence about quality problems in current quantum software. The study characterizes observed defects and their quality impacts but does not define a general quality model.

Testing, debugging, and formal verification provide complementary techniques for obtaining evidence about particular quality concerns. Ramalho \emph{et al.}~\cite{leite2025testing} review a broad range of testing and debugging approaches for quantum programs, including coverage, property-based, mutation, metamorphic, search-based, and fuzz testing, quantum assertions, bug patterns and benchmarks, program analysis, and debugging techniques. Their roadmap also highlights quantum-specific difficulties caused by probabilistic behavior, noise, no-cloning, transpilation, scalability, and hybrid quantum-classical interfaces. Formal reasoning and verification have likewise received substantial attention~\cite{chareton2021formal,ying2012floyd}. These techniques provide methods and evidence for evaluating particular criteria, but they do not themselves constitute a general software quality model.

Our work builds on and complements these lines of research. Existing studies have identified individual quality concerns, introduced non-functional requirements, proposed metrics for particular software attributes and artifacts, provided empirical evidence about quality problems, and developed techniques for testing, debugging, and verification. However, these contributions remain distributed across different development activities, artifacts, quality concerns, and measurement objectives. The present work instead constructs a base quality model for quantum software using a common factor-criterion-metric structure. It defines a broad set of classical and quantum-specific quality factors and criteria, establishes their many-to-many relationships, provides representative candidate metrics, and supports context-specific tailoring and lifecycle use. Existing requirements, measurement, testing, debugging, verification, and empirical studies can therefore complement the model by helping instantiate metrics, evaluate particular criteria, and provide evidence for assessing the corresponding quality factors.

\section{Conclusion}
\label{sec:conclusion}

This paper presents a base quality model for quantum software, organized using a factor-criterion-metric structure. Starting from representative classical software quality models and standards, we constructed a consolidated factor set, introduced quantum-specific extensions where needed, defined a set of quality criteria and a many-to-many mapping of core factor-to-criterion relationships, discussed representative candidate metrics and the conditions needed to interpret them, and illustrated the context-specific tailoring and lifecycle use of the model.

The proposed model organizes quantum software quality around a common set of factors, criteria, and candidate metrics while allowing adaptation to different application domains, platforms, and organizational contexts. Its factor set, criterion definitions, mappings, and representative candidate metrics provide a basis for further evaluation and refinement through expert feedback, case studies, and empirical studies.

Future work should focus on evaluating the proposed model through expert feedback, case studies, and empirical studies, improving tool support for collecting and interpreting quality evidence, developing well-defined, reproducible, and comparable measurement and reporting conventions, and studying how tailored quality profiles can be constructed for different classes of quantum software. In particular, further work is needed to examine how the proposed candidate metrics can be instantiated, interpreted, and compared under different backends, toolchains, execution settings, and application domains.

\bibliographystyle{ACM-Reference-Format}
\bibliography{ref}

\appendix
\section{Reverse Criterion-to-Factor Index}

For convenience, Table~\ref{table:criterion-factor-index} provides a reverse criterion-to-factor index derived directly from Table~\ref{table:factor-criterion-mapping}. This appendix table is intended only as a lookup aid for readers. The core factor-to-criterion mappings are presented in Table~\ref{table:factor-criterion-mapping}.

\begin{table*}[th]
\centering
\caption{Reverse Criterion-to-Factor Index Derived from Table~\ref{table:factor-criterion-mapping}}
\label{table:criterion-factor-index}
\renewcommand\arraystretch{1.2}
\footnotesize
\begin{tabular}{|p{1.1cm}|p{2.9cm}|p{12.2cm}|}
\hline
\multicolumn{1}{|l|}{\bf Type} &
\multicolumn{1}{l|}{\bf Criterion} &
\multicolumn{1}{l|}{\bf Factors Supported in Table~\ref{table:factor-criterion-mapping}} \\
\hline

\multirow{23}{*}{\centering\bf \makecell[c]{Classical\\ (McCall)}}
& Traceability & Correctness; Functionality; Safety; Verifiability; Manageability; Documentation\\
\cline{2-3}
& Completeness & Correctness; Functionality\\
\cline{2-3}
& Consistency & Correctness; Reliability; Maintainability; Functionality; Understandability; Modifiability; Safety; Verifiability; Manageability\\
\cline{2-3}
& Accuracy & Correctness; Reliability \\
\cline{2-3}
& Error tolerance & Reliability; Survivability; Safety \\
\cline{2-3}
& Simplicity & Reliability; Maintainability; Testability; Understandability; Modifiability\\
\cline{2-3}
& Modularity & Maintainability; Flexibility; Testability; Portability; Reusability; Modifiability; Expandability\\
\cline{2-3}
& Generality & Flexibility; Reusability; Expandability\\
\cline{2-3}
& Expandability & Flexibility; Expandability\\
\cline{2-3}
& Instrumentation & Testability; Survivability; Verifiability; Manageability\\
\cline{2-3}
& Self-descriptiveness & Maintainability; Flexibility; Testability; Portability; Reusability; Understandability; Supportability; Modifiability; Verifiability; Manageability; Documentation\\
\cline{2-3}
& Execution efficiency & Efficiency\\
\cline{2-3}
& Storage efficiency & Efficiency\\
\cline{2-3}
& Access control & Integrity; Security\\
\cline{2-3}
& Access audit & Integrity; Security\\
\cline{2-3}
& Operability & Usability; Supportability; Survivability; Safety; Interaction Capability; Manageability\\
\cline{2-3}
& Training & Usability; Supportability; Interaction Capability; Documentation\\
\cline{2-3}
& Communicativeness & Usability; Supportability; Interaction Capability\\
\cline{2-3}
& \makecell[l]{Software system\\independence} & Portability; Reusability\\
\cline{2-3}
& Machine independence & Portability; Reusability\\
\cline{2-3}
& \makecell[l]{Communications\\commonality} & Interoperability; Compatibility\\
\cline{2-3}
& Data commonality & Interoperability; Compatibility\\
\cline{2-3}
& Conciseness & Maintainability; Modifiability\\
\hline

\multirow{17}{*}{\centering\bf \makecell[c]{Classical\\ (Ext.)}}
& Resource utilization & Efficiency\\
\cline{2-3}
& Readability & Maintainability; Understandability\\
\cline{2-3}
& Change locality & Maintainability; Modifiability; Expandability\\
\cline{2-3}
& Documentation adequacy & Usability; Maintainability; Supportability; Documentation\\
\cline{2-3}
& Tool support & Testability; Maintainability; Supportability\\
\cline{2-3}
& Configurability & Expandability; Manageability\\
\cline{2-3}
& Reproducibility & Verifiability\\
\cline{2-3}
& Recoverability & Survivability\\
\cline{2-3}
& Confidentiality & Security\\
\cline{2-3}
& Authenticity & Security \\
\cline{2-3}
& Vulnerability resistance & Security\\
\cline{2-3}
& Hazard control & Safety\\
\cline{2-3}
& Fail-safe behavior & Safety\\
\cline{2-3}
& Interface conformance & Interoperability; Compatibility\\
\cline{2-3}
& Version compatibility & Portability; Compatibility\\
\cline{2-3}
& Result transparency & Usability; Verifiability; Interaction Capability; Documentation\\
\cline{2-3}
& Co-existence & Compatibility \\
\hline

\multirow{16}{*}{\centering\bf Quantum}
& Entanglement stability 
& Entanglement Robustness\\
\cline{2-3}
& \makecell[l]{Entanglement backend\\robustness} 
& Entanglement Robustness\\
\cline{2-3}
& \makecell[l]{Entanglement compilation\\robustness} 
& Entanglement Robustness\\
\cline{2-3}
& Backend portability 
& Portability; Quantum Hardware Adaptability\\
\cline{2-3}
& Calibration robustness 
& Reliability; Quantum Hardware Adaptability \\
\cline{2-3}
& Constraint tolerance 
& Quantum Hardware Adaptability\\
\cline{2-3}
& Orchestration correctness
& Correctness; Hybrid Quantum-Classical Interoperability \\
\cline{2-3}
& Synchronization correctness 
& Hybrid Quantum-Classical Interoperability\\
\cline{2-3}
& Orchestration robustness
& Reliability; Survivability; Hybrid Quantum-Classical Interoperability\\
\cline{2-3}
& Measurement bias 
& Correctness; Reliability \\
\cline{2-3}
& Statistical repeatability
& Reliability \\
\cline{2-3}
& Mitigation sensitivity 
& Reliability; Verifiability \\
\cline{2-3}
& Unitarity preservation 
& Physical Validity\\
\cline{2-3}
& \makecell[l]{Measurement\\semantics soundness} 
& Physical Validity\\
\cline{2-3}
& No-cloning compliance 
& Physical Validity\\
\cline{2-3}
& Assumption preservation
& Verifiability; Physical Validity \\
\hline
\end{tabular}
\end{table*}

\end{document}